\documentclass[journal]{IEEEtran}

\usepackage{graphicx}
\usepackage{multirow}
\usepackage{wrapfig}
\usepackage{subfigure}
\usepackage{amsmath}
\usepackage{cases}
\usepackage{array}
\usepackage{tikz}
\usepackage{tabularx}

\ifCLASSINFOpdf
\else
\fi

\begin{document}

\title{Noisy Speech Based Temporal Decomposition to Improve Fundamental Frequency Estimation}

\author{A.~Queiroz~and~R.~Coelho
\thanks{This work was supported in part by the National Council for Scientific and Technological Development (CNPq) 308155/2019 and Fundação de Amparo à Pesquisa do Estado do Rio de Janeiro (FAPERJ) under Grant 203075/2016 and in part by the Coordenação de Aperfeiçoamento de Pessoal de Nível Superior - Brasil (CAPES) - under Grant Code 001.}
\thanks{The authors are with the Laboratory of Acoustic Signal Processing, Military Institute of Engineering (IME), Rio de Janeiro, RJ 22290-270, Brazil (e-mail: coelho@ime.eb.br).}
}

\markboth{}%
{Shell \MakeLowercase{A. Queiroz and R. Coelho}: Noisy Speech Based Temporal Decomposition to Improve Fundamental Frequency Estimation}

\maketitle

\begin{abstract}

This paper introduces a novel method to separate noisy speech into low or high frequency frames, in order to improve fundamental frequency (F0) estimation accuracy. In this proposal, the target signal is analyzed by means of the ensemble empirical mode decomposition. Next, the pitch information is extracted from the first decomposition modes. This feature indicates the frequency region where the F0 of speech should be located, thus separating the frames into low-frequency (LF) or high-frequency (HF). The separation is applied to correct candidates extracted from a conventional fundamental frequency detection method, and hence improving the accuracy of F0 estimate. The proposed method is evaluated in experiments with CSTR and TIMIT databases, considering six acoustic noises under various signal-to-noise ratios. A pitch enhancement algorithm is adopted as baseline in the evaluation analysis considering three conventional estimators. Results show that the proposed method outperforms the competing strategies, in terms of low/high frequency separation accuracy. Moreover, the performance metrics of the F0 estimation techniques show that the novel solution is able to better improve F0 detection accuracy when compared to competitive approaches under different noisy conditions.

\end{abstract}


\IEEEpeerreviewmaketitle

\section{Introduction}

\IEEEPARstart{U}{rban} noisy acoustic scenarios may affect speech temporal and spectral attributes. A robust estimation of the fundamental frequency (F0) feature is a requirement for a diversity applications, such as speech coding \cite{GEURTS_2001}, speech synthesis \cite{ROSS_1999}, speaker and speech \cite{SAMBUR_1975}, \cite{LJOLJ_2002} recognition. In a voiced speech segment, the F0 consists on the vibration rate of the vocal folds, which corresponds to the inverse of the pitch period. The investigation of harmonic noisy components of speech signals has also gained significant attraction for strategies that attain intelligibility gain \cite{EALEY_2001}, \cite{WANG_2017}, \cite{QUEIROZ_2021}. These harmonic components and also the formants play a significant role for speech intelligibility in noise \cite{ERIKSSON_1995}, \cite{BROWN_2010}, \cite{WANG_2018}. Therefore, methods and systems which estimate fundamental frequency accurately need to be explored, particularly, at low signal-to-noise ratios (SNRs), where noise components may cause errors in estimation algorithms.

Several approaches have been proposed in the literature for F0 detection in the clean speech. Classical time domain methods are generally based on the Auto-Correlation Function (ACF) \cite{RABINER_1976}, \cite{ACF}. YIN \cite{YIN} is an alternative solution that uses the local minima of the Normalized Mean Difference Function (NMDF) with some post-processing procedures to avoid estimation errors caused by signal amplitude changes. Nevertheless, SHR \cite{SHR_2002} and SWIPE \cite{SWIPE} are examples of spectral techniques. The SHR method introduces the concept of sub-harmonic to harmonic ratio, and F0 detection is performed by looking for values which maximize this ratio. SWIPE estimates the F0 as the frequency of the sawtooth waveform whose spectrum best matches the spectrum of the input signal. Despite the variety of proposed methods, an accurate estimation of F0 in severe noisy conditions is still a challenging task in speech signal processing. 

In recent years, solutions have been proposed for F0 estimation in noisy speech \cite{PEFAC_2011}, \cite{BANA_2014}, \cite{YEG}. In \cite{PEFAC_2011}, the Pitch Estimation Filter with Amplitude Compression (PEFAC) is introduced, which applies a prefiltering to reduce the noise effects with interesting accuracy results. Machine learning based approaches \cite{LIU_2017}, \cite{DRUGMAN_2018}, \cite{CREPE_2018}, \cite{SPICE_2020} investigate classifiers that include feed-forward, recurrent and convolutional neural networks. 
Furthermore, other F0 detection methods \cite{TAO} \cite{HHT} are based on the Hilbert-Huang Transform (HHT) \cite{EMD_ORIG}. Specifically, the Empirical Mode Decomposition (EMD) \cite{EMD_ORIG} or its variations, which realizes a time-frequency (TF) decomposition to analyze the noisy speech signal for various tasks \cite{ZAO_2014}, \cite{COELHO_2015}, \cite{STALLONE_2020}. In the HHT-Amp technique \cite{HHT} the F0 estimates are achieved from the instantaneous amplitude functions of the target signal. Although these F0 estimation methods are examined for noisy environments, some errors can still occur, e.g., F0 subharmonics detection, octave errors, and indistinct periodic background noise from the voiced speech \cite{RAPT_1995}.

In a recent work \cite{GAZOR_2020}, a strategy is introduced to improve pitch detection attained by conventional estimation algorithms considering noisy scenarios. To this end, a Deep Convolutional Neural Network (DCNN) is trained in order to classify the voiced speech frames into low or high frequency. According to the classification, new F0 candidates are computed considering typical probable types of errors for a certain F0 value estimated by a detection method. Taking into account a procedure which involves two spectral attributes of speech frames and a cost function, the enhanced pitch is selected from the new candidates.


\begin{figure*}[t!]
\centering
\includegraphics[width=0.9\linewidth,keepaspectratio=true]{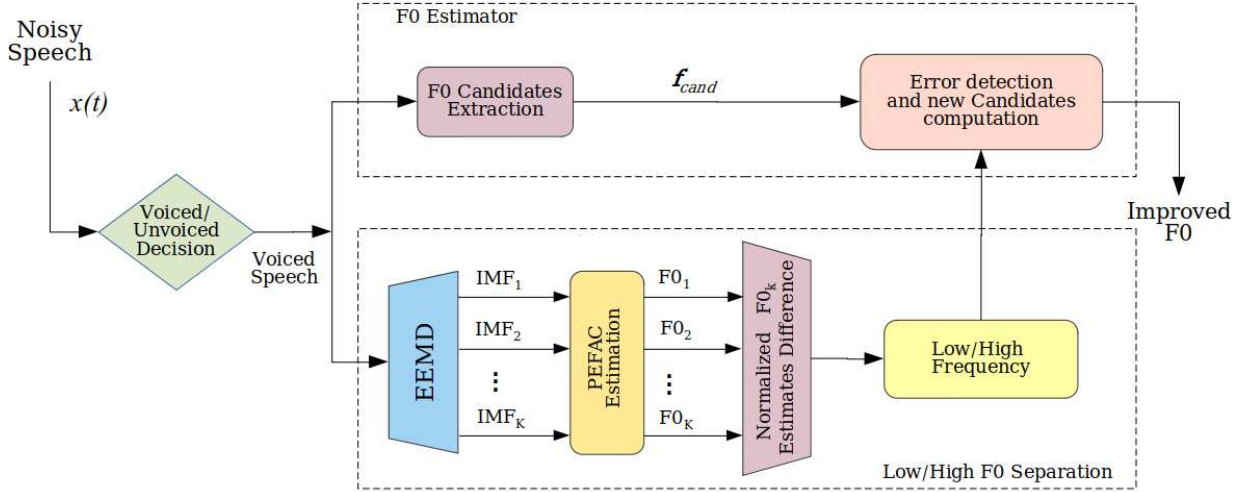}
\caption{Block diagram of the proposed method for low/high frequency separation and improvement of F0 estimation accuracy.}
\label{esquematico}
\end{figure*}


This work introduces a method (PRO) to separate harmonic noisy speech into low-frequency or high-frequency frames, in order to improve the F0 estimation accuracy. The proposed method applies the time-frequency Ensemble EMD (EEMD) \cite{EEMD_FLANDRIN} to decompose the noisy signal into a series of Intrinsic Mode Functions (IMFs). The first IMFs present the fastest oscillations referring to speech, which allows to attenuate the low-frequency noisy masking components. In this first step, PEFAC \cite{PEFAC_2011} is considered to estimate F0 from decomposition modes, expressing the low/high frequency tendency of speech frames. Then, a normalized distance that reflects the variation property of F0 is computed, selecting only the two IMFs with less variation in comparison to each other. Finally, the mean F0 of selected IMFs is compared to a threshold to indicate if the frame is considered low- or high-frequency. According the proposed separation, frequency candidates attained from a F0 detection method are corrected, leading to a set of enhanced candidates, and consequently improving the accuracy of F0 estimation.

Several experiments are conducted to examine the effectiveness and accuracy of the PRO method. For this purpose, speech utterances collected from two databases (CSTR \cite{CSTR} and TIMIT \cite{TIMIT_1993}) are corrupted by six real acoustic noises, considering five SNR values: -15 dB, -10 dB, -5 dB, 0 dB and 5 dB. The PRO method and DCNN approach \cite{GAZOR_2020} are examined in terms of improving the accuracy of fundamental frequency estimation, considering three F0 detection techniques: SHR \cite{SHR_2002}, SWIPE \cite{SWIPE} and HHT-Amp \cite{HHT}. The Gross Error rate (GE) \cite{GE_1976} and Mean Absolute Error (MAE) \cite{MAE_1985} are considered to evaluate the proposed and baseline methods. Experiments demonstrate that F0 approaches enhanced by PRO method achieve the lowest error values. Furthermore, PRO + HHT-Amp shows the best overall scores when compared to competitive techniques.

The main contributions of this work are:

\begin{itemize}
  
 \item Introduction of the PRO method for voiced speech frames separation into low/high frequency.
 
 \item Definition of a criterion to correct the candidates achieved from F0 estimation techniques.
 
 \item Accuracy improvement of F0 estimation approaches with error reduction from noisy speech signals with low SNR values.

\end{itemize}
 
 The remaining of this paper is organized as follows. Section II introduces the PRO method for voiced speech low/high frequency separation and correction of F0 candidates. Baseline F0 detection schemes are described in Section III, which also includes the comparative DCNN for error correction of pitch estimates. Section IV presents the evaluation experiments and results. Finally, Section V concludes this work.

\section{The Proposed Method}

The PRO method to improve the fundamental frequency estimation accuracy includes four main steps: time-frequency noisy speech signal decomposition (EEMD), F0 estimation of Intrinsic Mode Functions (IMFs), normalized distance computation, that reflects the F0 variation property for frame separation in low/high frequency, and finally, candidates correction from the F0 estimators. Fig. \ref{esquematico} illustrates the block diagram of the proposed method.

\subsection{TF Decomposition of Noisy Speech}

The first step is devoted to the Ensemble Empirical Mode Decomposition (EEMD). The general idea of the EMD \cite{EMD_ORIG} is to analyze a signal $x(t)$ between two consecutive extrema (minima or maxima), and define a local high-frequency part, also called detail $d(t)$, and a local trend $a(t)$, such that $x(t) = d(t) + a(t)$. An oscillatory IMF is derived from the detail function $d(t)$. The high- versus low-frequency separation procedure is iteratively repeated over the residual $a(t)$, leading to a new detail and a new residual. Thus, the decomposition leads to a series of IMFs and a residual, such that

\begin{equation}
 x\left ( t \right ) = \sum_{k=1}^{K} \mathrm{IMF}_k \left ( t \right ) + r\left ( t \right )
\end{equation}
where $\mathrm{IMF}_k \left ( t \right )$ is the $k$-th mode of $x(t)$ and $r(t)$ is the residual. In contrast to other signal decomposition methods, a set of basis functions is not demanded for the EMD. Besides, this strategy leads to fully data-driven decomposition modes and does not require the stationarity of the target signal.

The EEMD was introduced in \cite{EEMD} to overcome the mode mixing problem that generally occurs in the original EMD \cite{EMD_ORIG}. The key idea is to average IMFs obtained after corrupting the original signal using several realizations of White Gaussian Noise (WGN). Thus, EEMD algorithm can be described as:

\begin{enumerate}
 \item Generate $x^n\left ( t \right ) = x\left ( t \right ) + w^n\left ( t \right )$ where $w^n\left ( t \right )$, $n = 1, \cdots, N$, are different realizations of WGN;
 
 \item Apply EMD to decompose $x^n\left ( t \right )$, $n = 1, \cdots, N$, into a
series of components $\mathrm{IMF}_k^n \left ( t \right )$, $k = 1, \cdots, K$;

\item Assign the $k$-th mode of $x(t)$ as

\begin{equation}
 \mathrm{IMF}_k \left ( t \right ) = \frac{1}{N} \sum_{n=1}^{N} \mathrm{IMF}_k^n \left ( t \right );
\end{equation}

\item Finally, $x\left ( t \right ) = \sum_{k=1}^{K} \mathrm{IMF}_k \left ( t \right ) + r\left ( t \right )$, where $r(t)$ is the residual.

\end{enumerate}

Although the lack of mathematical formulation, it is interesting to point out that EEMD is a very powerful tool for analyzing non-stationary real signals and has been successfully applied in several research areas \cite{STALLONE_2020}, \cite{HUANG_2014}.

\subsection{F0 Estimation of IMFs}

\begin{figure}[t!]
 \centering
 \includegraphics[width=0.7\linewidth,keepaspectratio=true]{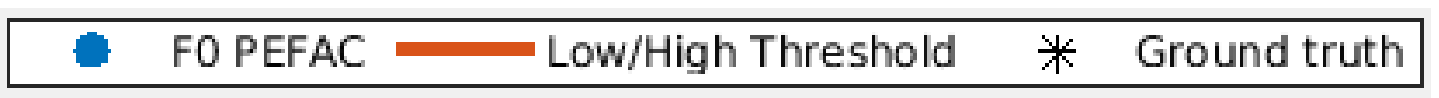}
 \label{leg_pef}
 \vspace{-0.3cm}
\end{figure}

\begin{figure}[t!]
\centering
 \subfigure[]{
  \includegraphics[width=0.49\linewidth,keepaspectratio=true, clip=true,trim=0pt 0pt 0pt 0pt]{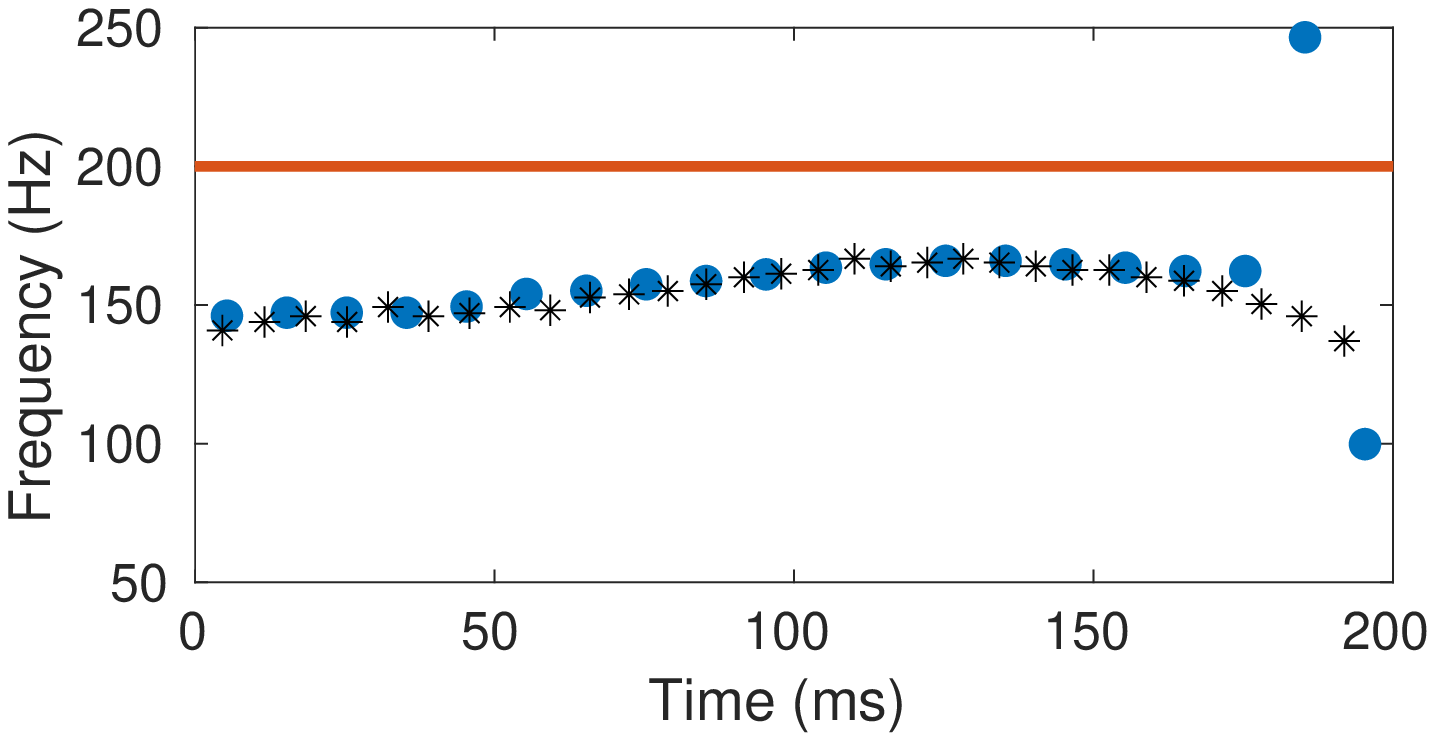}
}
\hspace{-0.6cm}
 \subfigure[]{
  \includegraphics[width=0.49\linewidth,keepaspectratio=true, clip=true,trim=0pt 0pt 0pt 0pt]{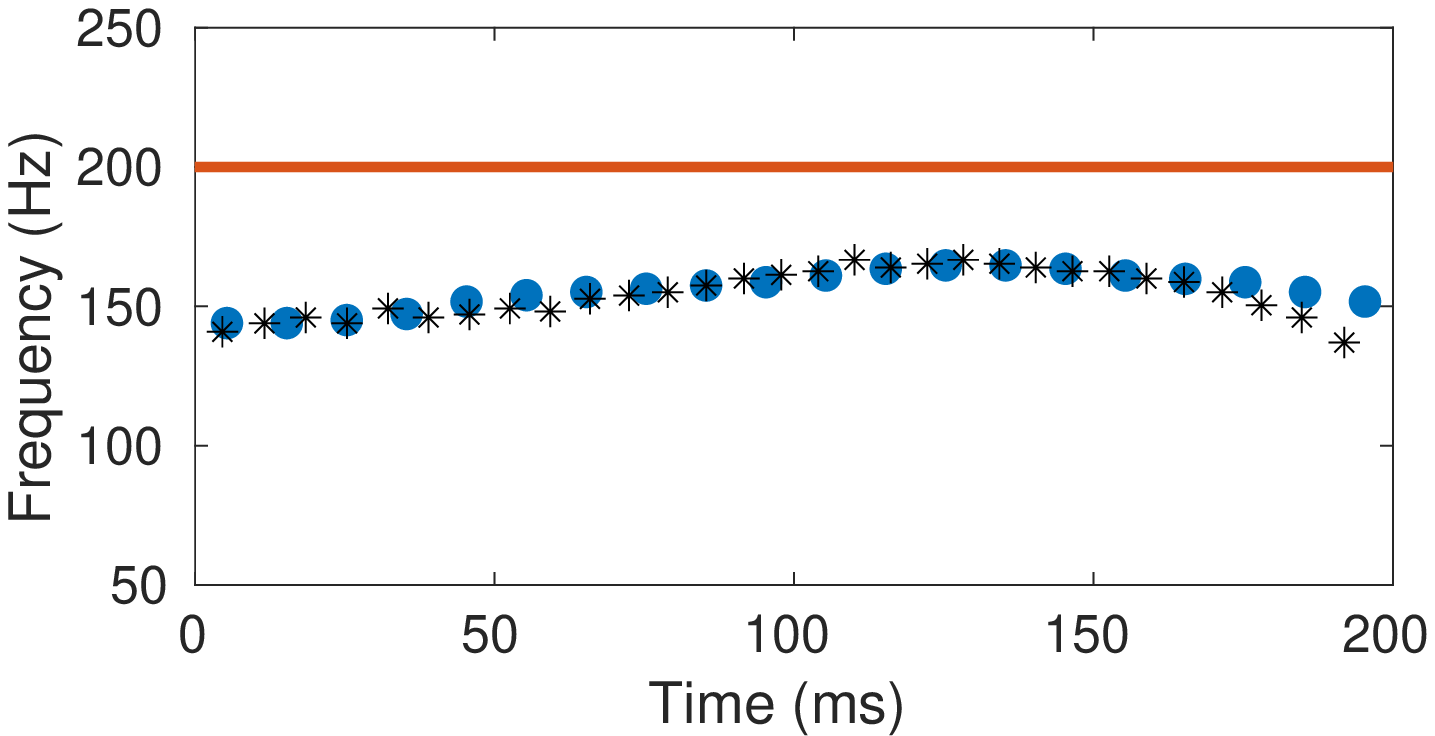}
}
\hspace{-0.01cm}
  \subfigure[]{
  \includegraphics[width=0.49\linewidth,keepaspectratio=true, clip=true,trim=0pt 0pt 0pt 0pt]{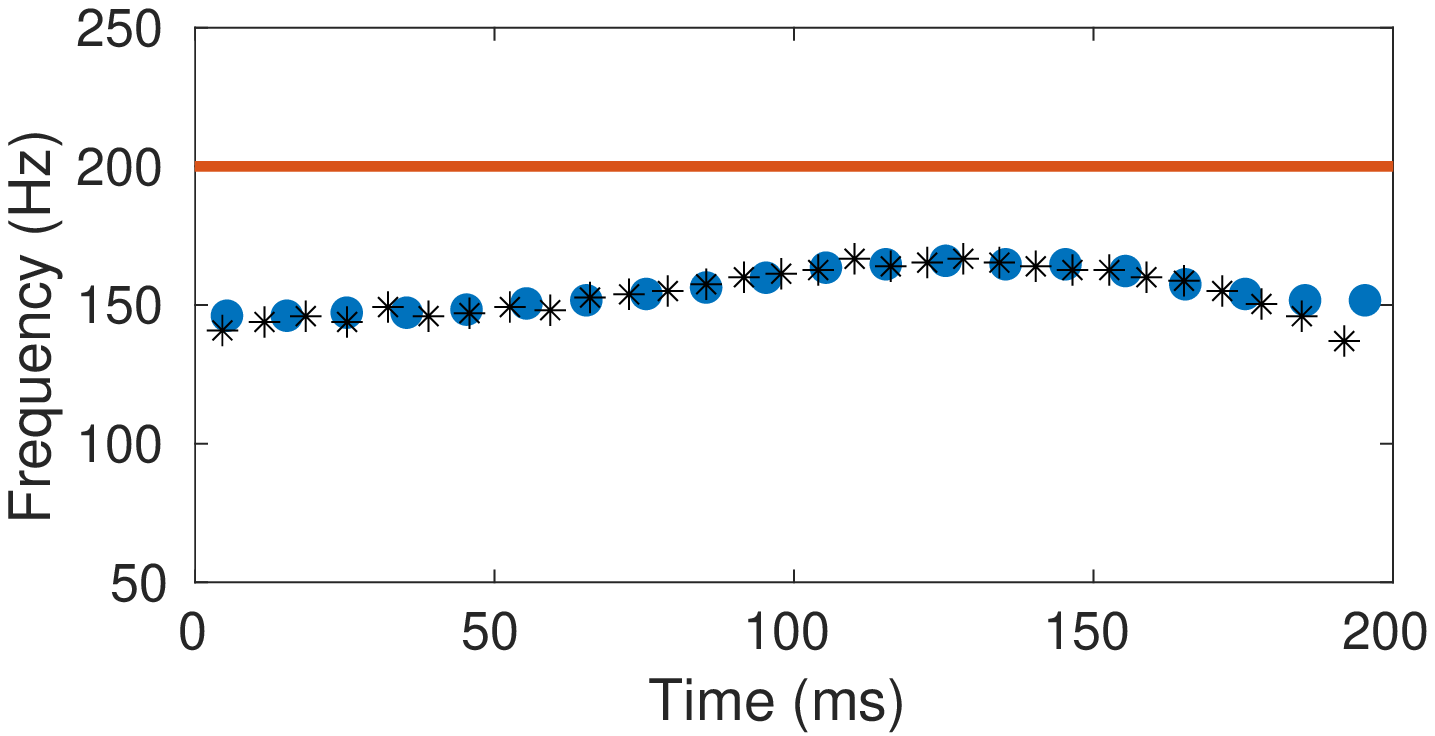}
}
\hspace{-0.6cm}
 \subfigure[]{
  \includegraphics[width=0.49\linewidth,keepaspectratio=true, clip=true,trim=0pt 0pt 0pt 0pt]{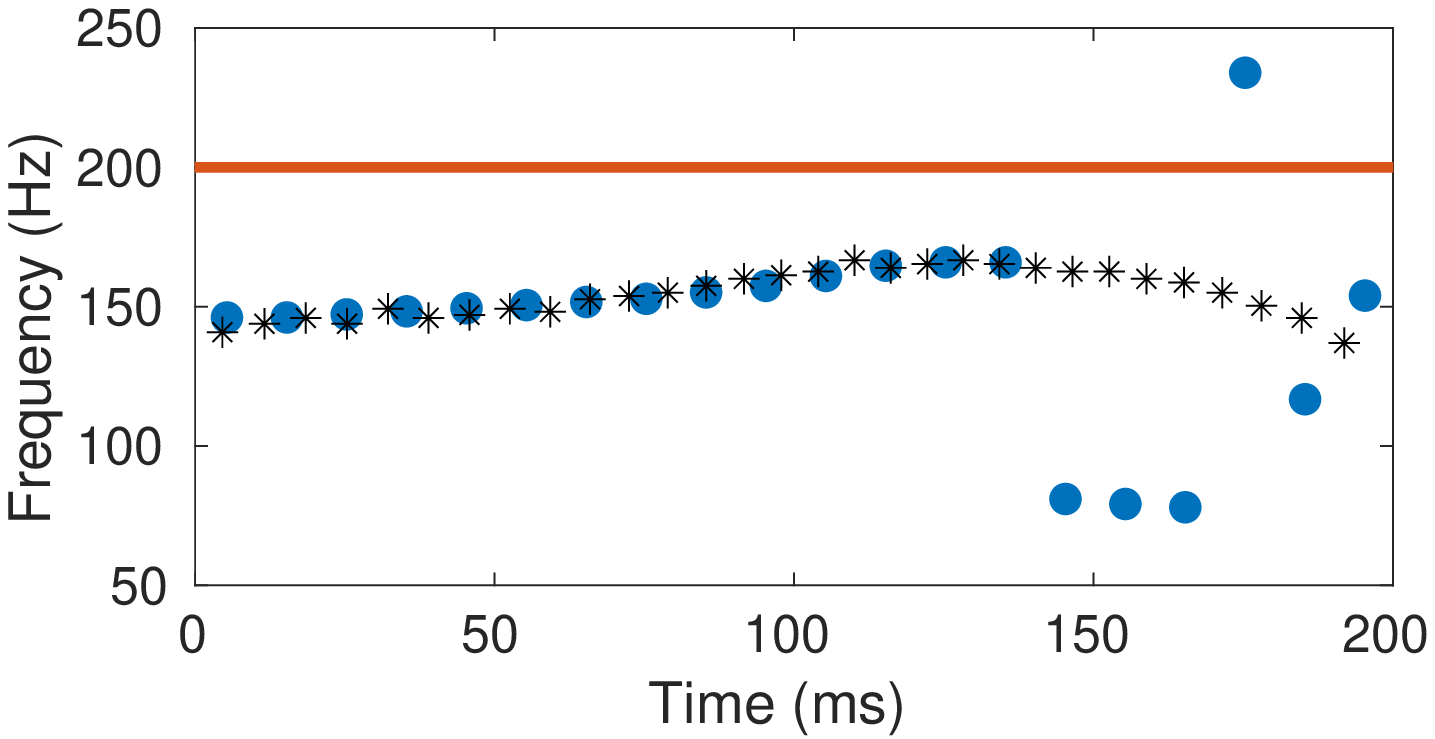}
}
 \caption{F0 estimation with PEFAC for (a) IMF$_\text{1}$, (b) IMF$_\text{2}$, (c) IMF$_\text{3}$ and (d) IMF$_\text{4}$ of a speech segment corrupted with Babble noise and SNR = 0 dB.}
\label{pefac_est}

\end{figure}

In this step, the F0 of frames of each IMF is estimated by the selected PEFAC \cite{PEFAC_2011} algorithm. This solution detects the F0 by convolving the power spectrum of the frame in the log-frequency domain, using a filter that sums the energy of the pitch harmonics. This filter highlights the harmonic components, improving the noise-robustness of the algorithm. Thus, let $\hat{F}0_{k,q}$ denote the F0 value estimated from frame $q$ of $\mathrm{IMF}_k \left ( t \right )$, the 
$\hat{F}0_{q}$ vector is composed as

\begin{equation}
 \hat{F}0_{q} = \begin{bmatrix}
\hat{F}0_{1,q}\\ 
\hat{F}0_{2,q}\\ 
\vdots \\ 
\hat{F}0_{K,q}
\end{bmatrix},
\end{equation}
to express the tendency that the frame is placed in a low or high frequency region. Although the first IMFs are composed of fastest oscillations, these components also present some nonnegligible low-frequency content of the speech signal \cite{HHT}, \cite{MEDINA_2021}. Therefore, in this paper it is considered only the first four IMFs ($K=4$) in order to avoid the acoustic noise masking effect, whose energy is mostly concentrated at low frequencies \cite{ZAO_2014}, \cite{MEDINA_2021}, \cite{CHATLANI_2012}.

Fig. \ref{pefac_est} shows the F0 estimated with PEFAC for the first four IMFs of a voiced speech segment collected from the CSTR database \cite{CSTR} and corrupted by Babble \cite{RSG10} noise with SNR = 0 dB. Default parameters of PEFAC are considered, whose voiced segment is split into overlapping frames with 90 ms length and 10 ms frame shift. Note that for the analyzed IMFs, the F0 contours are close to the ground truth. In most frames the frequencies have similar values for all the IMFs, particularly when the decomposition attenuates the noisy component of speech. However, some noisy masking effect should appears on IMFs, leading to a difference in the F0, as seen in Fig. \ref{pefac_est}(a),(d) (IMF$_1$ and IMF$_4$) in the frames between 150-200 ms.

\subsection{Low/High Frequency Separation}

A normalized distance is computed between IMFs for the successive frames, in order to detect and overcome the differences in the estimated F0. Let $k$ and $k'$ denote IMF indexes, the distance $\delta^q_{\hat{F}0}$ is described as

\begin{equation}
 \delta^q_{\hat{F}0}(k,k') = \left | \frac{\hat{F}0_{k,q} - \hat{F}0_{k',q}}{\hat{F}0_{k,q} + \hat{F}0_{k',q}} \right |.
\end{equation}
This distance is determined for different values of $k$ and $k'$, resulting in the following matrix

\begin{equation}
 \delta^q_{\hat{F}0} = \begin{bmatrix}
0 & \delta^q_{\hat{F}0}\left ( 1,2 \right ) & \cdots  & \delta^q_{\hat{F}0}\left ( 1,K \right ) \\ 
\delta^q_{\hat{F}0}\left ( 2,1 \right ) & 0 & \cdots  & \delta^q_{\hat{F}0}\left ( 2,K \right ) \\ 
\vdots  & \vdots  & \ddots  & \vdots \\ 
\delta^q_{\hat{F}0}\left ( K,1 \right ) & \delta^q_{\hat{F}0}\left ( K,2 \right ) & \cdots  & 0
\end{bmatrix}.
\end{equation}
The row components of the matrix are summed, and the resulting values obtained express the variation property for the $k$-th IMF. The two IMFs with the smallest variation scores are selected, and the frequency region is defined as the mean value of PEFAC F0 estimates ($\bar{F}0_{q}$). Finally,  a low-frequency to high-frequency threshold $\gamma$ is adopted, and the separation is performed such as

\begin{equation} \label{medins}
\begin{cases}
 \bar{F}0_{q} \leq \gamma, &  \text{low-frequency frame};\\
\bar{F}0_{q} > \gamma, &  \text{high-frequency frame}.
\end{cases}
\end{equation}
In \cite{TITZE_1994} it is shown that the variability of speech F0 is between 50-200 Hz for men and 120-350 Hz for women. Therefore, in this study, a threshold of $\gamma$ = 200 Hz is considered, since this is an average value for both genders of speakers.

\subsection{Extraction and Correction of Candidates}

\begin{figure}[t!]
\centering
 \subfigure[]{
  \includegraphics[width=0.8\linewidth,keepaspectratio=true]{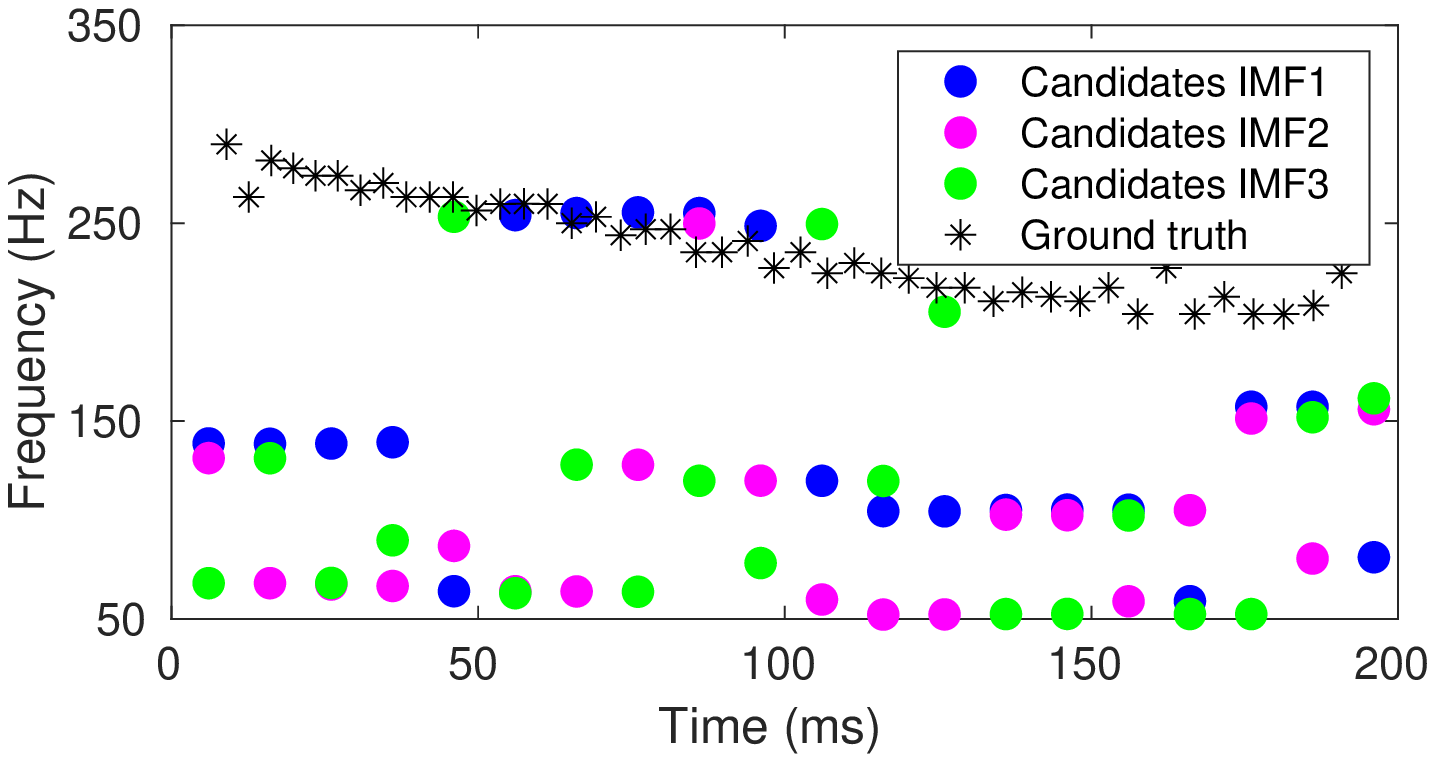}
}
 \subfigure[]{
  \includegraphics[width=0.8\linewidth,keepaspectratio=true, clip=true,trim=0pt 0pt 0pt 0pt]{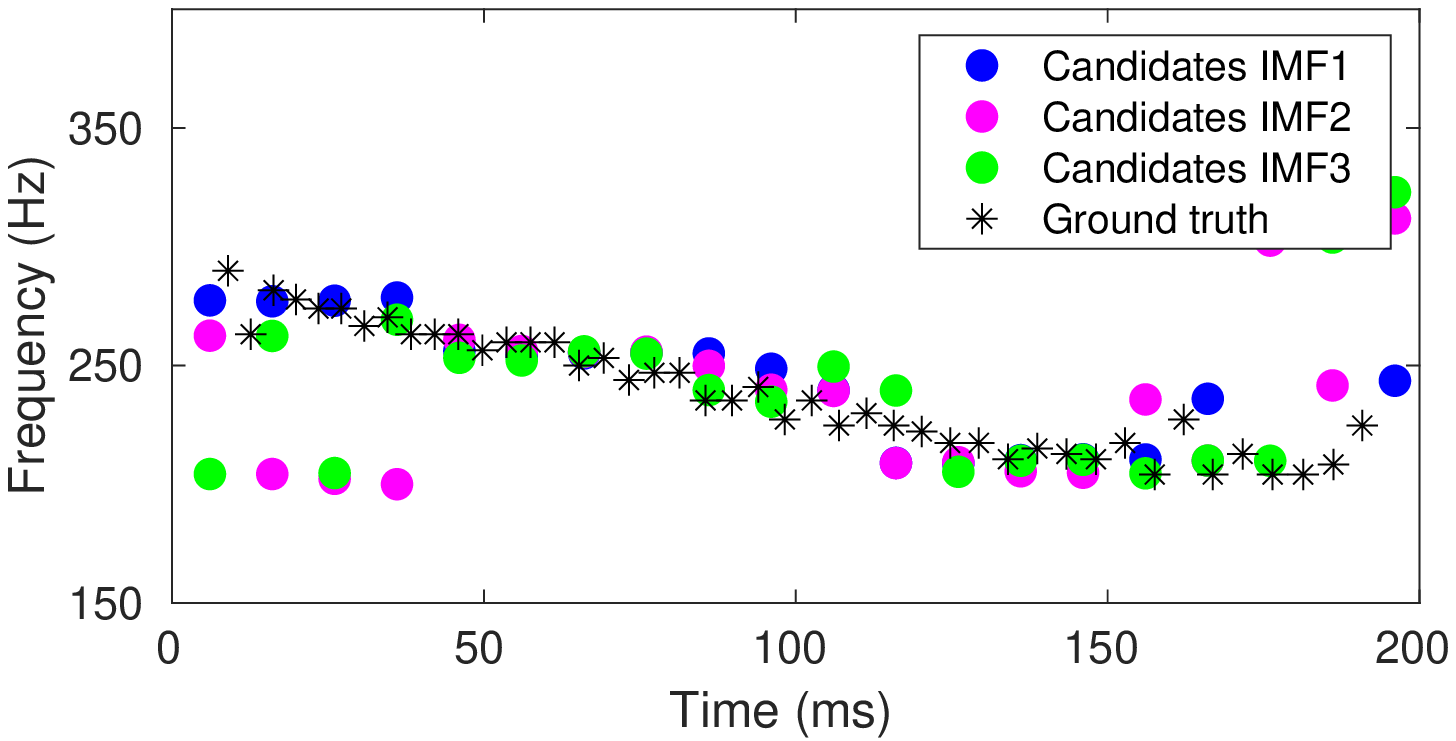}
}
 \caption{(a) Original F0 Candidates from HHT-Amp method for a high-frequency speech segment corrupted by Babble noise with SNR = 0 dB, and (b) candidates adjusted by the proposed separation and correction algorithm.}
\label{adjust_cands}

\end{figure}

The F0 estimation techniques are prone to different types of errors. Typical errors in F0 estimation, e.g., halving and doubling errors, are caused by the detection of harmonics other than the first, resulting in F0 estimates that are multiples of the true F0 \cite{RAPT_1995}. In order to overcome this issue, the proposed method aims to correct the frequency candidates extracted by a F0 detection algorithm according to these types of errors. The PRO method states that, the F0 candidates ($f_{cand}$) must lie in [50,200] Hz or [200,400] Hz, for low-frequency and high-frequency frames, respectively. Thus, considering a low-frequency frame, $f_{cand}$ is corrected following the criteria

\begin{equation}\label{criteriolow}
 f_c = \begin{cases}
f_{cand}, & 50 \leq f_{cand} \leq 200\\ 
0.5f_{cand}, & 200 <  f_{cand} \leq 400 \\ 
0.25f_{cand}, & f_{cand} > 400
\end{cases}.
\end{equation}
where $f_c$ is the corrected F0 candidate. Finally, the high-frequency frame is adjusted as follows:

\begin{equation}\label{criteriohigh}
 f_c = \begin{cases}
4f_{cand}, & 50 \leq f_{cand} \leq 100\\ 
2f_{cand}, & 100 <  f_{cand} \leq 200 \\ 
f_{cand}, & 200 <  f_{cand} \leq 400 \\
0.5f_{cand}, & f_{cand} > 400
\end{cases}.
\end{equation}

Fig. \ref{adjust_cands} illustrates the F0 candidates correction procedure for a speech segment corrupted by Babble noise with SNR = 0 dB, with 200 ms duration.  Fig. \ref{adjust_cands}(a) denotes the F0 candidates extracted using the HHT-Amp \cite{HHT} estimation technique, which computes a set of three candidates (one candidate from each of the first three IMFs) for a 10 ms time interval. Ground truth indicates that the entire speech segment is classified as high-frequency. Note that most candidates are located in the low-frequency region ($f_{cand} \leq$ 200 Hz). Fig. \ref{adjust_cands}(b) shows the new candidates, which are corrected according to the criteria described in (\ref{criteriolow}) and (\ref{criteriohigh}). Moreover, observe that the adjusted F0 candidates take place in high-frequency region, matching the ground truth.

\section{F0 Detection Baseline Techniques}

This Section briefly describes the DCNN based technique \cite{GAZOR_2020} adopted as baseline solution for the low/high frequency separation experiments. Further, the F0 estimation accuracy improvement is evaluated considering the SHR \cite{SHR_2002}, SWIPE \cite{SWIPE} and HHT-Amp \cite{HHT} approaches.

\subsection{DCNN low/high frequency Separation}

This technique \cite{GAZOR_2020} consists on training a DCNN, in order to classify a voiced speech frame into low ($F0 \leq 200$ Hz) or high ($F0 > 200$ Hz) frequency. The DCNN architecture adopted is based on VGGNet \cite{VGGNET_2015}, with six convolutional layers, three Fully-Connected (FC) layers and an output classification layer (Softmax). The DCNN input sequence is a 60 ms extracted directly from the voiced speech samples with a sampling rate of 16 kHz, or 960 samples.

According the low- or high-frequency classification, new F0 candidates $f_j$ are extracted based on initial estimation attained by a conventional method $f_p$. The improved F0 estimate value is selected from the set of new candidates by exploiting a restrained selection procedure. Two spectral attributes are introduced to assist in selecting the improved F0 among candidates: Weighted Euclidean Deviation ($d_{j,q}$) and Weighted Comb Filtering ($y_{j,q}$). The first attribute is computed observing the first five frequency peaks positions in each frame, resulting in a peak vector ${P}_{i,q}=[P_{1,q}, P_{2,q}, \dots, P_{5,q}]$. For a pitch candidate $f_{j,q}$, a candidate vector is defined as $V_{j,q}=[v_{j1,q}, v_{j2,q}, \dots, v_{j5,q}]$, where $v_{ji,q}$ is the point-wise ratio between the peak vector and the candidate $f_{j,q}$. Thus, the Weighted Euclidean Deviation is described as 

\begin{equation}
 d_{j,q}=\left \| \left (V_j - T  \right ) \odot U  \right \|_2,
\end{equation}
where $T = [1,2,3,4,5]$ denotes the multiple harmonics of pitch, $U_i = 1/T_i$ and $\odot$ is the point-wise multiplication. The Weighted Comb Filtering for $f_{j,q}$ is defined as

\begin{equation}
 y_{j,q} = \sum_{f} X_q\left ( f \right )C\left ( f/f_{j,q} \right ),
\end{equation}
where $X_q\left ( f \right )$ is the power spectrum for the $q$th frame, and $C\left ( f/f_{j,q} \right )=\left ( \frac{1}{2}+\frac{1}{2}\mbox{cos}\left ( 2\pi f/f_{j,q} \right )\mbox{exp}\left ( -f/f_{j,q} \right ) \right )$. Finally, a cost function is defined as
\begin{equation}
\mbox{cost}_q = \left | \mbox{log}f_{j,q} - \mbox{log}f_{i,q+1} \right | + \frac{\lambda }{pr_q\left ( \frac{y_{j,q}}{\alpha } + \frac{1}{d_{j,q}} + \varepsilon_q  \right )},
\hspace{-0.05cm}
\end{equation}
where $\left | \mbox{log}f_{j,q} - \mbox{log}f_{i,q+1} \right |$ is a F0 smoothness feature, $\lambda > 0$ and $\alpha$ are regularization parameters, $pr_q$ is the low/high frequency probability of the softmax layer of DCNN, and $\varepsilon_q =1$ if $f_j = f_p$, or zero otherwise. The smaller cost function value for the F0 candidates is more likely to be the true F0 value.

\subsection{SHR}

SHR method \cite{SHR_2002} is based on the definition of a parameter measure called Sub-Harmonic-Harmonic Ratio, designed to describe the amplitude ratio between subharmonics and harmonics. Let $A(f)$ denote the amplitude spectrum for each short-term signal, the sum of harmonic amplitude is described as

\begin{equation}
 SH = \sum_{n=1}^{N}A\left (nF0 \right),
\end{equation}
where $N$ and $F0$ are the maximum number of harmonics considered in the spectrum and fundamental frequency, respectively. In other hand, the sum of subharmonic amplitude is achieved assuming at one half of F0

\begin{equation}
 SS = \sum_{n=1}^{N}A\left ( \left ( n-\frac{1}{2} \right ) F0 \right ).
\end{equation}
Finally, $SHR$ is the ratio between $SS$ and $SH$:

\begin{equation}
 SHR = \frac{SS}{SH}.
\end{equation}
In \cite{SHR_2002}, this ratio is computed on a logarithmic frequency scale. If the obtained $SHR$ value is greater than a certain threshold, subharmonics frequencies are considered in the analysis. Otherwise, the harmonic frequencies are adopted in the F0 detection.

\subsection{SWIPE}

The core idea of SWIPE \cite{SWIPE} is that if a signal is periodic with fundamental frequency $f$, its spectrum must contain peaks at multiples of $f$ and valleys in between. Since each peak is surrounded by two valleys, the Average Peak-to-Valley Distance (APVD) for the $n$-th peak is defined as
\vspace{0.1cm}
\begin{equation}
\hspace{-0.1cm}\scalebox{0.93}{
 $d_n(f) = \left | X(nf) \right | - \frac{1}{2} \left [ \left | X\left ( \left ( n-\frac{1}{2} \right ) f \right ) \right | 
 + \left | X\left ( \left ( n+\frac{1}{2} \right ) f \right ) \right | \right ]$
 }\hspace{-0.5cm}
\end{equation}
where $X$ is the estimated spectrum of the signal which takes frequency as input and outputs corresponding density. The global APVD is achieved by averaging over the first $p$ peaks

\begin{equation}
 D_n(f) = \frac{1}{p} \sum_{n=1}^{p} d_n(f).
\end{equation}
The F0 estimated is the $f$ value that maximizes this function, searching $f$ in the range [50 500]Hz candidates, with samples distributed every $1/48$ units on a base-2 logarithmic scale.

\subsection{HHT-Amp}

The HHT-Amp method is summarized as follows:
\begin{itemize}
\item Apply the EEMD as described in Section II-A to decompose the voiced sample sequence $x_q(t)$.

\item Compute the instantaneous amplitude functions $a_{k,q}(t) = |Z_{k,q}(t)|, k = 1, \ldots, K, $ from the analytic signals
defined as 
\begin{equation}
 Z_{k,q}(t) = \mbox{IMF}_{k,q}(t) + j \, H\{\mbox{IMF}_{k,q}(t)\},
\end{equation}
where $H\{\mbox{IMF}_{k,q}(t)\}$ refers to the Hilbert transform of $\mbox{IMF}_{k,q}(t)$.

\item Calculate the ACF $r_{k,q}(\tau) = \sum_t  a_k(t) \, a_k(t+\tau)$ of the amplitude functions $a_{k,q}(t), k = 1, \ldots, K$.

\item For each decomposition mode $k$, let $\tau_0$ be the lowest $\tau$ value that correspond to an ACF peak, subject to 
$\tau_{min} \leq \tau_0 \leq \tau_{max}$. The restriction is applied according to the range $[F_{min}, F_{max}]$
of possible $F_0$ values. The $k$-th pitch candidate is defined as $\tau_0 / {f_s}$, 
where $f_s$ refers to the sampling rate.

\item Apply the decision criterion defined in \cite{HHT} to select the best pitch candidate $\hat T_0$. The estimated $F_0$ is
given by $\hat F_0 = 1 / \hat T_0$.

\end{itemize}

In \cite{HHT}, it was shown that the HHT-Amp method achieves interesting results in estimating the fundamental frequency of noisy speech signals. HHT-Amp was evaluated in a wide range of noisy scenarios, including five acoustic noises, outperforming four competing estimators in terms of GE and MAE.


\begin{table*}[t!]
\begin{center}
\caption{Low/High Frequency Separation Errors (\%)}

{
\renewcommand{\arraystretch}{1.2}
\setlength{\tabcolsep}{3.pt}
\begin{tabular}{llccccccccccccccccccccccccc}
\hline
&\multirow{2}{*}&\multicolumn{5}{c}{Babble}&&\multicolumn{5}{c}{Cafeteria}&&\multicolumn{5}{c}{SSN}&&\multicolumn{5}{c}{Volvo}&&\\
\cline{3-7}\cline{9-13}\cline{15-19}\cline{21-25}

&SNR(dB)&-15&-10&-5&0&5&&-15&-10&-5&0&5&&-15&-10&-5&0&5&&-15&-10&-5&0&5&&Avg.\\\hline

\multirow{2}{*}{CSTR}&DCNN&65.4&55.8&49.7&38.6&25.4&&68.2&54.2&44.1&30.7&20.5&&65.8&59.3&54.5&40.0&24.1&&34.8&23.7&18.1&16.1&14.5&&40.2 \\

&PRO&\textbf{39.7}&\textbf{25.1}&\textbf{14.5}&\textbf{7.7}&\textbf{5.0}&&\textbf{36.2}&\textbf{25.0}&\textbf{13.8}&\textbf{7.8}&\textbf{4.9}&&\textbf{42.9}&\textbf{31.0}&\textbf{15.7}&\textbf{7.6}&\textbf{3.8}&&\textbf{2.5}&\textbf{2.8}&\textbf{2.6}&\textbf{2.9}&\textbf{3.0}&&\textbf{14.7}\\\hline

\multirow{2}{*}{TIMIT}&DCNN&26.2&19.6&17.3&15.3&14.0&&26.3&19.9&17.5&15.2&14.0&&26.3&20.1&17.7&15.4&14.3&&18.9&15.5&13.7&12.7&12.6&&17.0\\

&PRO&\textbf{26.1}&\textbf{18.5}&\textbf{12.7}&\textbf{8.4}&\textbf{6.2}&&\textbf{25.2}&\textbf{18.8}&\textbf{13.0}&\textbf{8.9}&\textbf{6.2}&&\textbf{23.7}&\textbf{17.0}&\textbf{10.5}&\textbf{7.5}&\textbf{5.6}&&\textbf{5.4}&\textbf{5.5}&\textbf{4.7}&\textbf{4.0}&\textbf{3.7}&&\textbf{11.6}\\\hline

\end{tabular}\label{classification_acc}
}
\end{center}

\end{table*}


\section{Results and Discussion}

This Section presents the accuracy results attained by PRO method for low/high frequency separation, in comparison to DCNN based technique. Following, GE and MAE metrics are adopted to evaluate the accuracy for the competitive approaches in F0 estimation improvement experiments, considering several noisy environments.

\subsection{Speech and Noise Databases}

The experiments consider the CSTR \cite{CSTR} and a subset of TIMIT \cite{TIMIT_1993} databases to evaluate the competitive methods. CSTR is composed of 100 English utterances spoken by male (50) and female (50) speakers, sampled at 20 KHz. The reference F0 values are available based on the recordings of laryngograph data. The TIMIT subset is composed of 128 speech signals spoken by 8 male and 8 female speakers, sampled at 16 KHz and with 3 s average duration. The reference F0 values are obtained from \cite{GONZALEZ_2014}. 

Six noises are used to corrupt the
speech utterances: acoustic Babble and Volvo attained from RSG-10 \cite{RSG10}, Cafeteria, Train and Helicopter from Freesound.org\footnote{[Online]. Available: https://freesound.org.}, and Speech Shaped Noise (SSN) from DEMAND \cite{DEMAND_2013} database. Experiments are conducted considering noisy speech signals with five SNR values.

\subsection{Evaluation Metrics}

For the evaluation, it is adopted the Gross Error rate (GE) and Mean Absolute Error (MAE). GE \cite{GE_1976} is defined as:

\begin{equation}
 GE = \frac{P_{error}}{P} \times 100,
\end{equation}
where $P$ denotes the total number of voiced frames, and $P_{error}$ is the number of voiced frames for which the deviation estimated F0 from the ground truth is more than 20\%. MAE \cite{MAE_1985} is computed as

\begin{equation}
 MAE = \frac{1}{Q} \sum_{q=1}^{Q} \left | \hat{F}0\left ( q \right ) - F0 \left ( q \right ) \right |,
\end{equation}
where $Q$ is the total number of frames, $\hat{F}0\left ( q \right )$ the estimate and $F0 \left ( q \right )$ is the reference. This metric provides a greater perception of the error, since it indicates an absolute distance (in Hz) between F0 reference and estimation.


\begin{figure*}[t!]
\hspace{0.5cm}
\centering
  \includegraphics[width=0.92\linewidth,keepaspectratio=true]{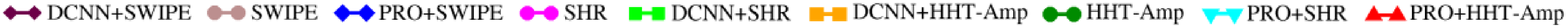}
  \vspace{-0.4cm}
\end{figure*}

\begin{figure*}[t!]
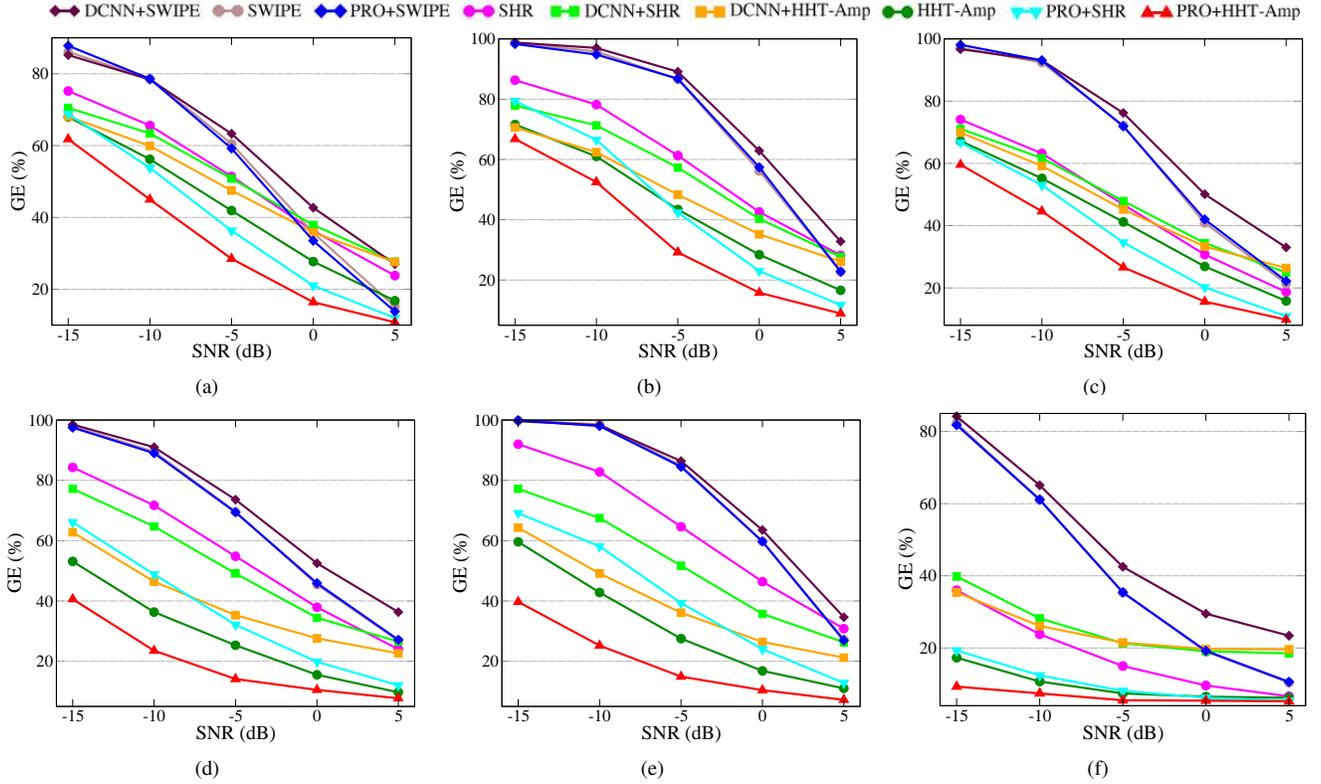

\centering
 \subfigure[]{
  \includegraphics[width=0.295\linewidth,keepaspectratio=true]{GE_Babble_CSTR.eps}
}
\hspace{-0.01cm}
 \subfigure[]{
  \includegraphics[width=0.3\linewidth,keepaspectratio=true]{GE_SSN_CSTR.eps}
}
\hspace{-0.01cm}
  \subfigure[]{
  \includegraphics[width=0.3\linewidth,keepaspectratio=true]{GE_Cafeteria_CSTR.eps}
}
\hspace{-0.01cm}
 \subfigure[]{
  \includegraphics[width=0.3\linewidth,keepaspectratio=true]{GE_Train_CSTR.eps}
}
\hspace{-0.01cm}
 \subfigure[]{
  \includegraphics[width=0.3\linewidth,keepaspectratio=true]{GE_Helicopter_CSTR.eps}
}
\hspace{-0.01cm}
 \subfigure[]{
  \includegraphics[width=0.3\linewidth,keepaspectratio=true]{GE_Volvo_CSTR.eps}
}
 \caption{The GE averaged over CSTR database speech signals versus five SNR values, for six noisy conditions: (a) Babble, (b) SSN, (c) Cafeteria, (d) Train, (e) Helicopter, and (f) Volvo. Detected F0 improved considering the low/high frequency separation errors.}
\label{GE_Ruidos_CSTR}

\end{figure*}

\subsection{Low/High Frequency Separation Accuracy}

Table \ref{classification_acc} presents the error results with the PRO and the competing DCNN technique for low/high frequency separation of voiced speech frames. These results denotes the mean values for the utterances of CSTR and TIMIT databases, respectively, considering the speech signals corrupted by four noises and five SNR values. Note that PRO outperforms the DCNN for all of the noisy scenarios. The proposed method achieves the lowest error values in all the 20 noisy conditions and the two databases. For instance, it can be seen that PRO attains interesting results for the most severely conditions, for SNR = -15 dB. In this case, the error scores are about 28 percentage points (p.p.) smaller than those achieved by DCNN approach in CSTR database. The overall errors obtained 
with PRO are 14.7\% and 11.6\% for CSTR and TIMIT databases, against 40.2\% and 17.0\% for DCNN, respectively. 

The interesting accuracy results attained by PRO method are particularly important, and can be justified by the fact that it is considered only the first four IMFs of EEMD. This attenuates the noise noise masking effects, since the most part of noisy energy is concentrated in low frequencies, e.g., IMF$_\text{6}$, IMF$_\text{7}$ and IMF$_\text{8}$. Moreover, PEFAC also contributes to the separation accuracy, due to the filter introduced in the algorithm, which rejects high-level narrow-band noise and favors the correct F0 estimates, even in severe noisy environments.

\subsection{GE and MAE Results}

The proposed method and DCNN baseline technique for improvement of F0 detection accuracy are compared in terms of GE and MAE, considering three F0 estimation approaches: SHR, SWIPE and HHT-Amp. In this study, it is assumed that errors in separation of the voiced speech into low-frequency and high-frequency signals can generate some error into the whole system. Furthermore, voiced/unvoiced detection has been done by the strategy based on the Zero Crossing (ZC) rate and energy of the speech signal as described in \cite{BACHU_2008}.


\begin{table}[t!] \caption{\label{GE_TIMIT} Gross Error (GE) results (\%) with the Proposed and Baseline methods for TIMIT database.}

\renewcommand{\arraystretch}{1.4}
\setlength{\tabcolsep}{3.pt}
\begin{center}
{
\begin{tabular}{crccccccc}

\hline
&&\multicolumn{3}{c}{DCNN}&&\multicolumn{3}{c}{PRO}\\
\cline{3-5}\cline{7-9}
Noise&\multicolumn{1}{c}{SNR}&SHR&\scriptsize{SWIPE}&\scriptsize{HHT}\tiny{-Amp}&&SHR&\scriptsize{SWIPE}&\scriptsize{HHT}\tiny{-Amp}\\\hline

\multirow{6}{*}{Babble} &-15 dB&71.2&85.9&65.9&&69.5&84.7&\textbf{57.3}\\
&-10 dB&66.3&83.7&57.9&&60.3&81.7&\textbf{44.1}\\
&-5 dB&57.0&78.5&46.0&&46.7&74.9&\textbf{29.5}\\
&0 dB&45.9&64.6&36.1&&32.4&56.5&\textbf{16.3}\\
&5 dB&36.6&48.0&28.8&&20.7&34.8&\textbf{9.4}\\
&Average&55.4&72.2&46.9&&45.9&66.5&\textbf{31.3}\\ \hline

\multirow{6}{*}{SSN} &-15 dB&82.1&98.9&70.1&&81.9&98.0&\textbf{64.4}\\
&-10 dB&78.9&98.7&60.6&&71.8&97.6&\textbf{48.5}\\
&-5 dB&67.5&97.0&47.5&&53.2&94.8&\textbf{27.6}\\
&0 dB&52.8&89.1&35.6&&34.4&83.4&\textbf{15.5}\\
&5 dB&40.2&71.8&28.1&&21.1&59.3&\textbf{8.3}\\
&Average&64.3&91.1&48.4&&52.5&86.6&\textbf{32.9}\\ \hline

\multirow{6}{*}{Cafeteria} &-15 dB&70.9&97.4&64.5&&66.9&98.7&\textbf{56.8}\\
&-10 dB&65.7&97.3&56.1&&57.1&97.8&\textbf{39.8}\\
&-5 dB&56.5&93.1&45.5&&43.4&92.0&\textbf{25.0}\\
&0 dB&44.5&81.0&35.1&&29.1&73.9&\textbf{15.3}\\
&5 dB&35.0&62.8&27.7&&18.4&49.8&\textbf{8.4}\\
&Average&54.5&86.3&45.8&&43.0&82.4&\textbf{29.1}\\ \hline

\multirow{6}{*}{Train} &-15 dB&82.3&99.4&56.2&&70.4&99.0&\textbf{37.9}\\
&-10 dB&72.6&96.1&46.5&&56.3&94.8&\textbf{24.9}\\
&-5 dB&60.6&89.5&36.4&&42.3&84.1&\textbf{14.5}\\
&0 dB&48.7&79.7&29.9&&30.1&71.4&\textbf{10.3}\\
&5 dB&38.9&67.7&26.3&&20.2&56.2&\textbf{8.1}\\
&Average&60.6&86.5&39.0&&43.9&81.1&\textbf{19.1}\\ \hline

\multirow{6}{*}{Helicopter} &-15 dB&86.0&99.8&63.3&&71.4&99.9&\textbf{42.9}\\
&-10 dB&80.5&99.5&49.5&&63.0&99.9&\textbf{27.2}\\
&-5 dB&67.0&97.4&37.6&&48.4&97.3&\textbf{17.1}\\
&0 dB&53.3&90.6&29.7&&32.7&87.3&\textbf{9.7}\\
&5 dB&41.9&76.9&24.5&&20.8&65.0&\textbf{7.1}\\
&Average&65.7&92.8&40.9&&47.3&89.9&\textbf{20.8}\\ \hline

\multirow{6}{*}{Volvo} &-15 dB&49.3&95.7&33.9&&30.9&95.1&\textbf{7.9}\\
&-10 dB&39.7&90.7&27.8&&20.7&84.7&\textbf{7.0}\\
&-5 dB&31.9&78.0&24.1&&13.5&65.9&\textbf{5.9}\\
&0 dB&27.0&58.7&22.2&&9.2&42.4&\textbf{4.8}\\
&5 dB&24.1&44.2&21.8&&6.6&26.2&\textbf{4.3}\\
&Average&34.4&73.5&26.0&&16.2&62.8&\textbf{6.0}\\ \hline

\multicolumn{2}{c}{Overall}&55.8&83.7&41.2&&41.4&78.2&\textbf{23.2}\\\hline
\end{tabular}
}
\end{center}
\vspace{-0.3cm}
\end{table}


Fig. \ref{GE_Ruidos_CSTR} depicts the GE values for the F0 estimation approaches improved by PRO method, in comparison with the performance of DCNN based competitive techniques, averaged over signals of CSTR database. It is interesting to mention that PRO + HHT-Amp achieves the lowest error results, being the most accurate method when compared to baseline solutions. In addition, note that PRO method attains interesting F0 improvement for SHR and HHT-Amp F0 estimators, even assuming the low/high frequency separation errors. For instance, the original HHT-Amp approach attained a GE of 59.6\% for Helicopter noise (Fig. \ref{GE_Ruidos_CSTR}(e)) with SNR = -15 dB, in contrast with 39.7\% of PRO + HHT-Amp scheme. For the SHR method, DCNN improved the F0 estimates in some cases, e.g., Train and Helicopter noises, but it is still surpassed by PRO. However, for SWIPE and HHT-Amp, the errors of DCNN separation affect whole system, causing an increasing in the GE scores. Note that, in severe noisy condition, the SWIPE method obtained the highest GE values. This fact may be explained by its inner voiced/unvoiced detection that is impaired by the background situation.

Table \ref{GE_TIMIT} presents GE results of PRO and DCNN fundamental frequency improvement methods obtained with speech signals of TIMIT database. Note that PRO outperforms DCNN for SHR and HHT-Amp detectors. For example, considering the Helicopter noise with SNR = -15 dB, the GE rate for the SHR F0 estimation scheme decreased from 86.0\% with the DCNN \cite{GAZOR_2020} to 71.4\% with PRO, i.e., a reduction of 14.6 p.p. Likewise the results exposed for CSTR, PRO + HHT-Amp achieves the best accuracy results for TIMIT database, with lowest GE values in all the 30 noisy conditions. For the Babble noise, this strategy attains the average GE of 31.3\% against 46.9\% for PRO + SHR. On overall average, PRO + HHT-Amp presents a GE score of 23.2\%, which is 18.0 p.p., 18.2 p.p and 32.6 p.p smaller than DCNN + HHT-Amp, PRO + SHR and DCNN + SHR baseline approaches, respectively.


\begin{figure*}[t!]
\centering
 \subfigure[]{
  \includegraphics[width=0.31\linewidth,keepaspectratio=true,clip=true,trim=0pt 0pt 0pt 15pt]{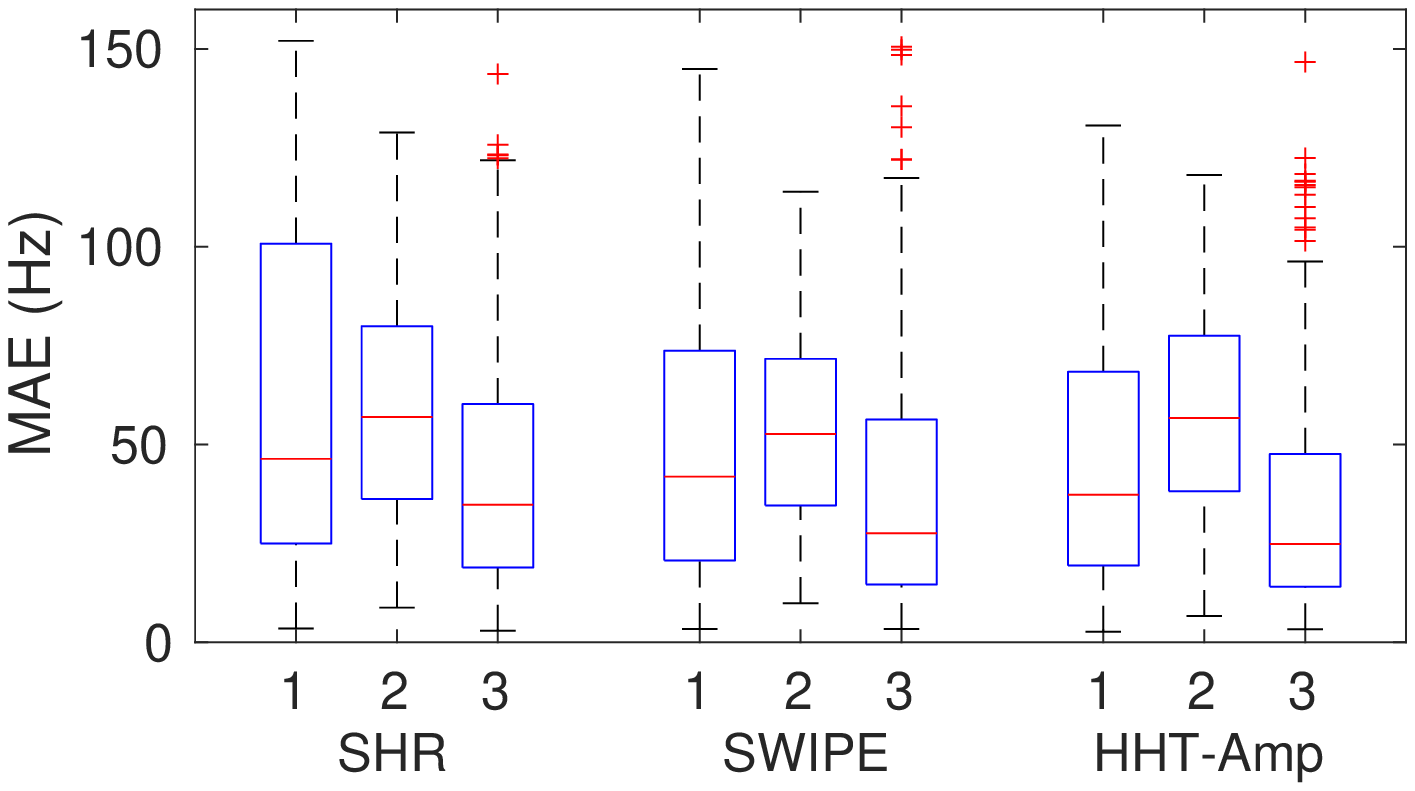}
}
\hspace{-0.3cm}
 \subfigure[]{
  \includegraphics[width=0.31\linewidth,keepaspectratio=true,clip=true,trim=0pt 0pt 0pt 15pt]{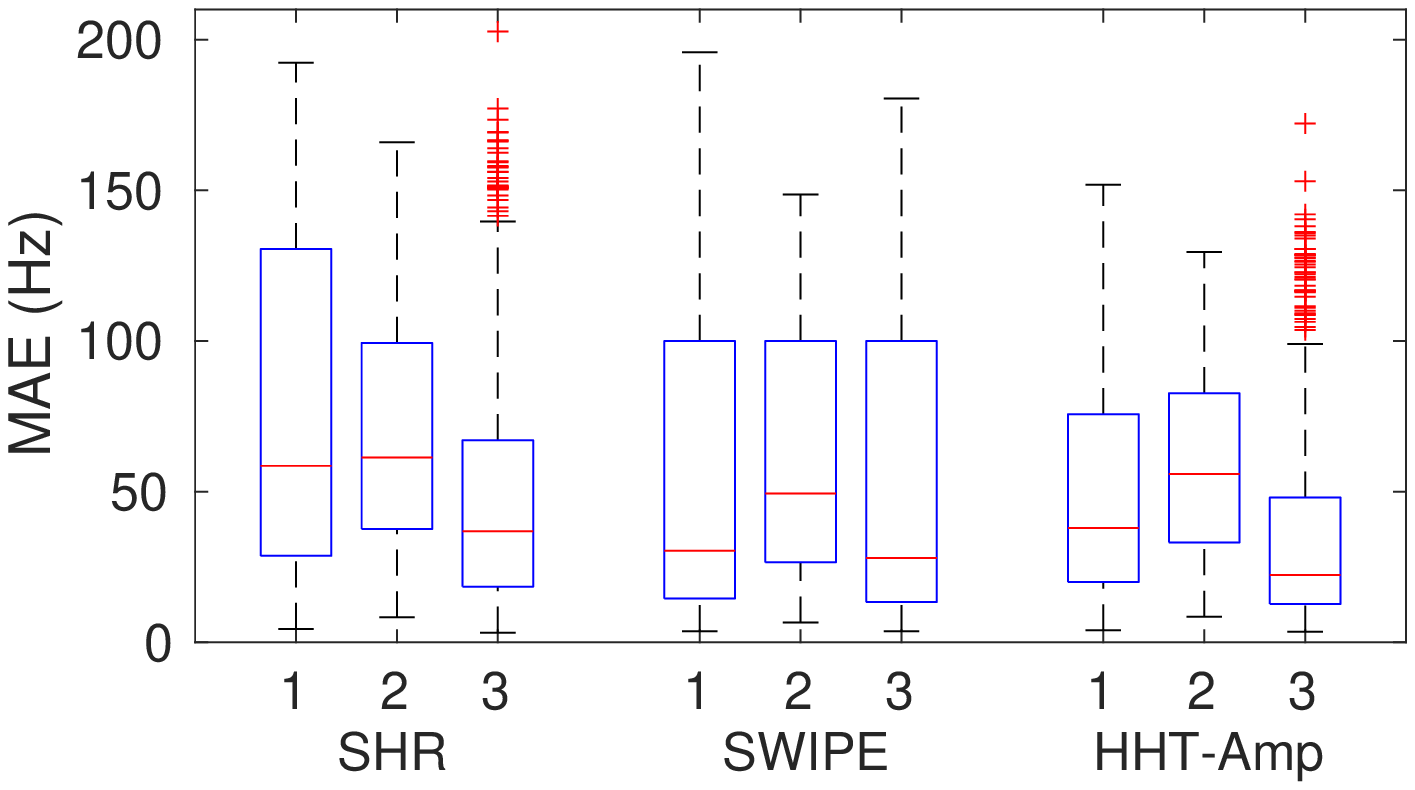}
}
\hspace{-0.3cm}
  \subfigure[]{
  \includegraphics[width=0.31\linewidth,keepaspectratio=true,clip=true,trim=0pt 0pt 0pt 15pt]{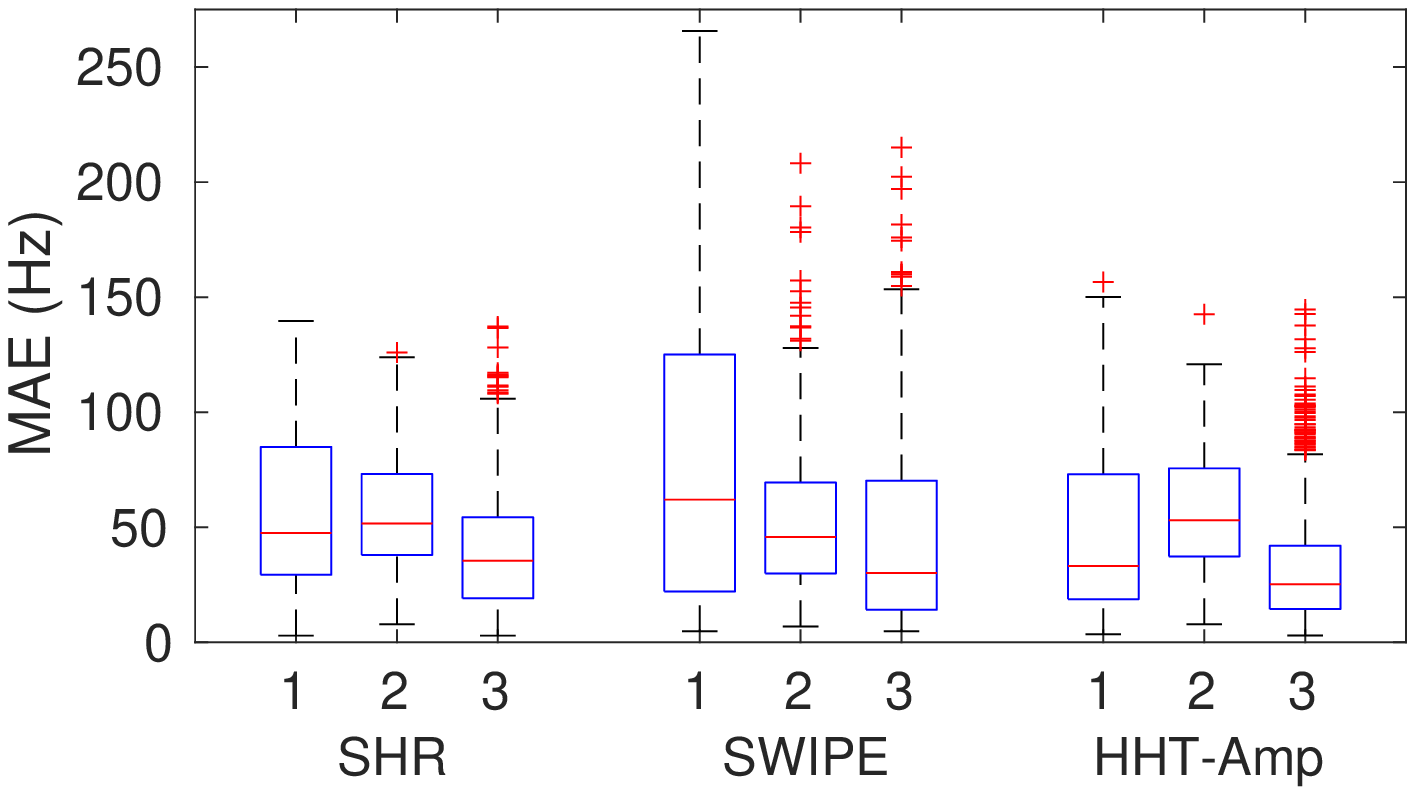}
}
\hspace{-0.01cm}
 \subfigure[]{
  \includegraphics[width=0.31\linewidth,keepaspectratio=true,clip=true,trim=0pt 0pt 0pt 15pt]{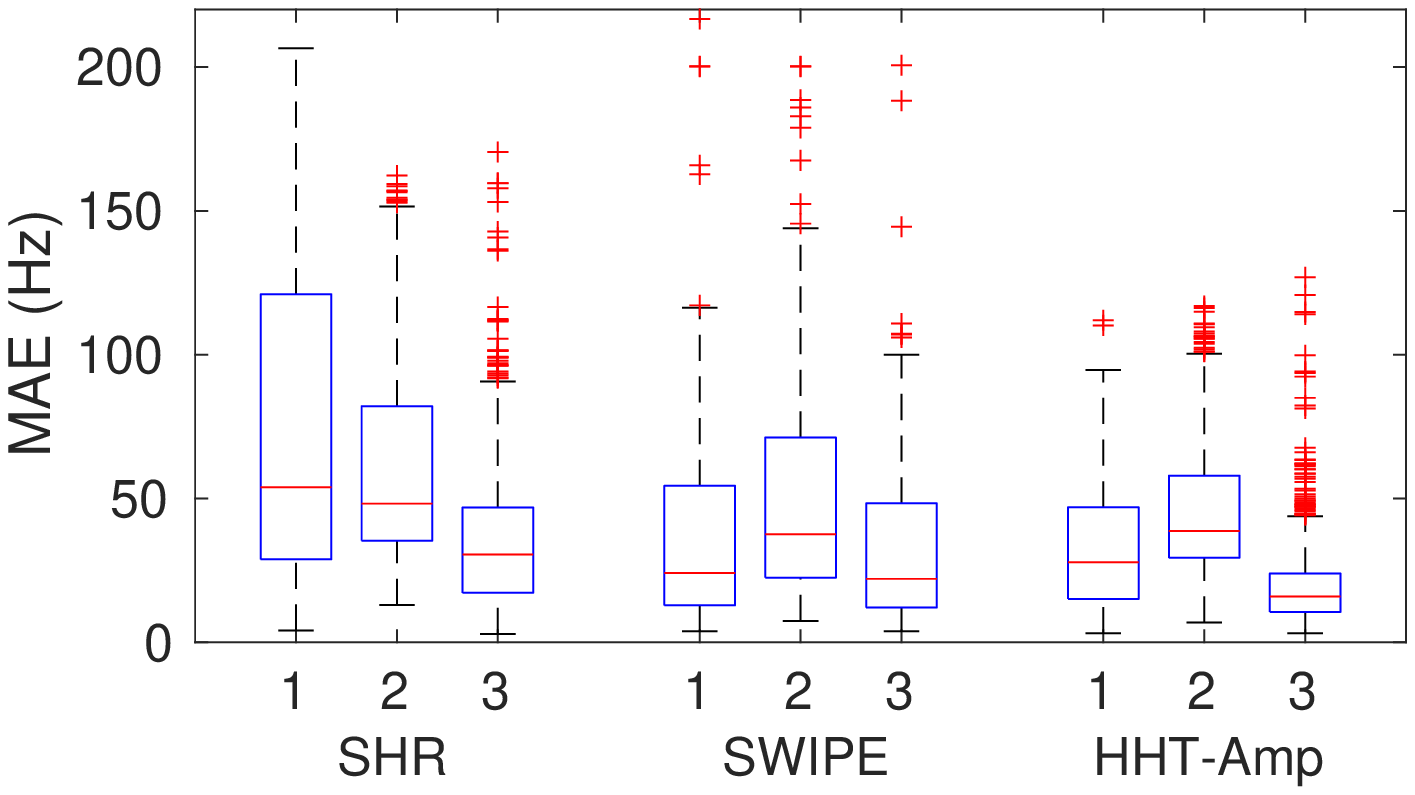}
}
\hspace{-0.3cm}
 \subfigure[]{
  \includegraphics[width=0.31\linewidth,keepaspectratio=true,clip=true,trim=0pt 0pt 0pt 15pt]{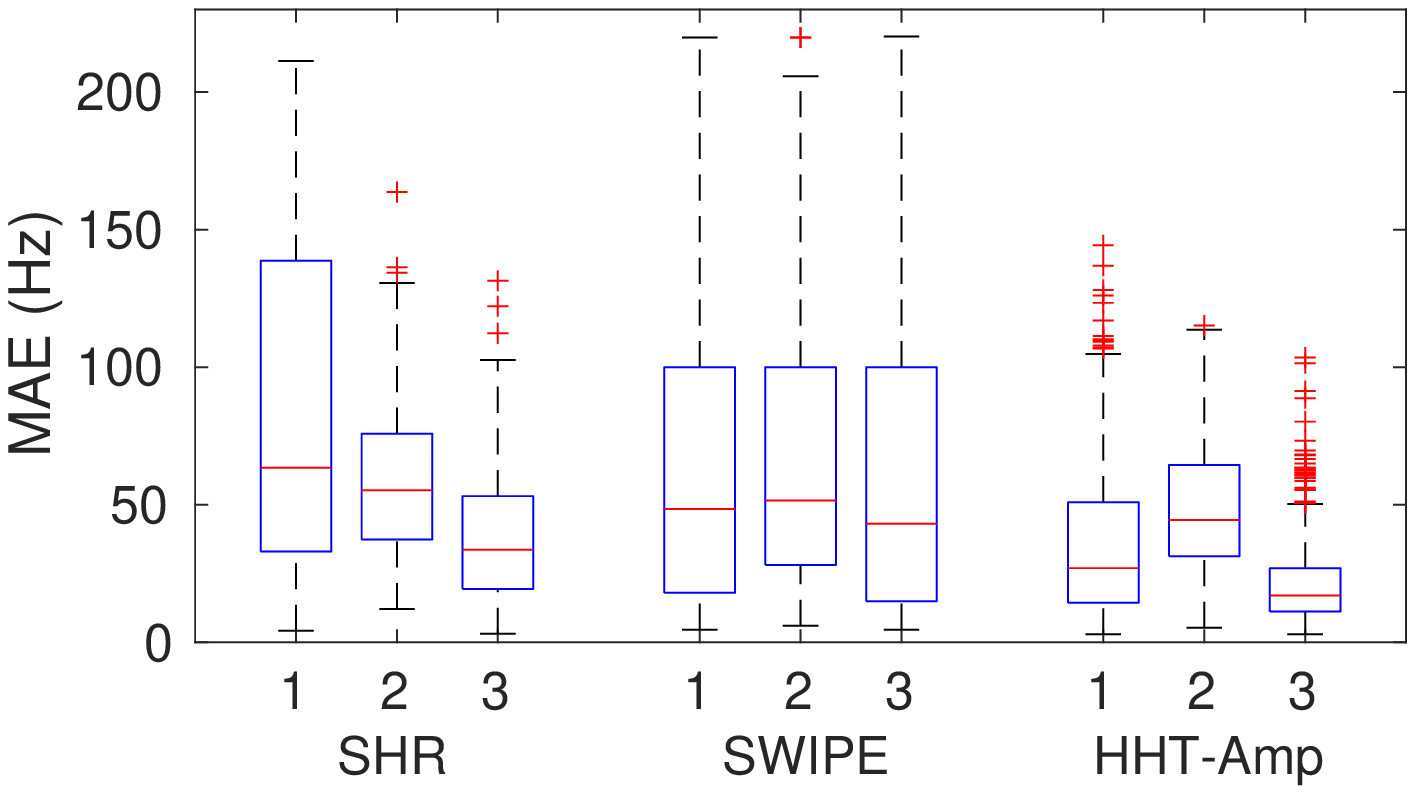}
}
\hspace{-0.3cm}
 \subfigure[]{
  \includegraphics[width=0.31\linewidth,keepaspectratio=true,clip=true,trim=0pt 0pt 0pt 15pt]{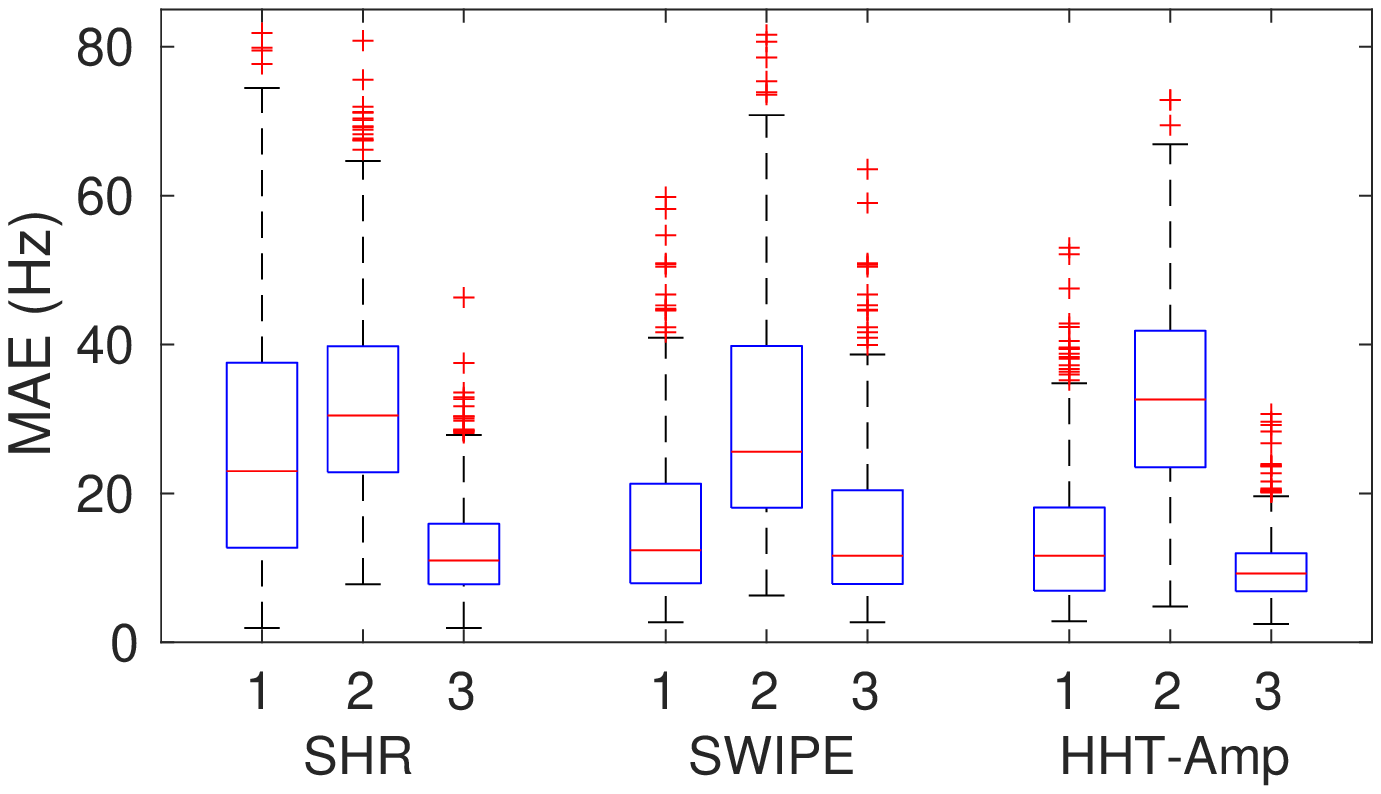}
}
 \caption{The MAE box-plots for CSTR database speech signals with five SNR values, for six noisy conditions: (a) Babble, (b) SSN, (c) Cafeteria, (d) Train, (e) Helicopter, and (f) Volvo. Case 1 refers to original F0 estimation method, case 2 the improved F0 by DCNN and case 3 improved F0 by PRO. Results obtained considering the low/high frequency separation errors.}
\label{MAE_Ruidos_CSTR}

\end{figure*}


\begin{figure*}[t!]
\centering
 \subfigure[]{
  \includegraphics[width=0.31\linewidth,keepaspectratio=true,clip=true,trim=0pt 0pt 0pt 15pt]{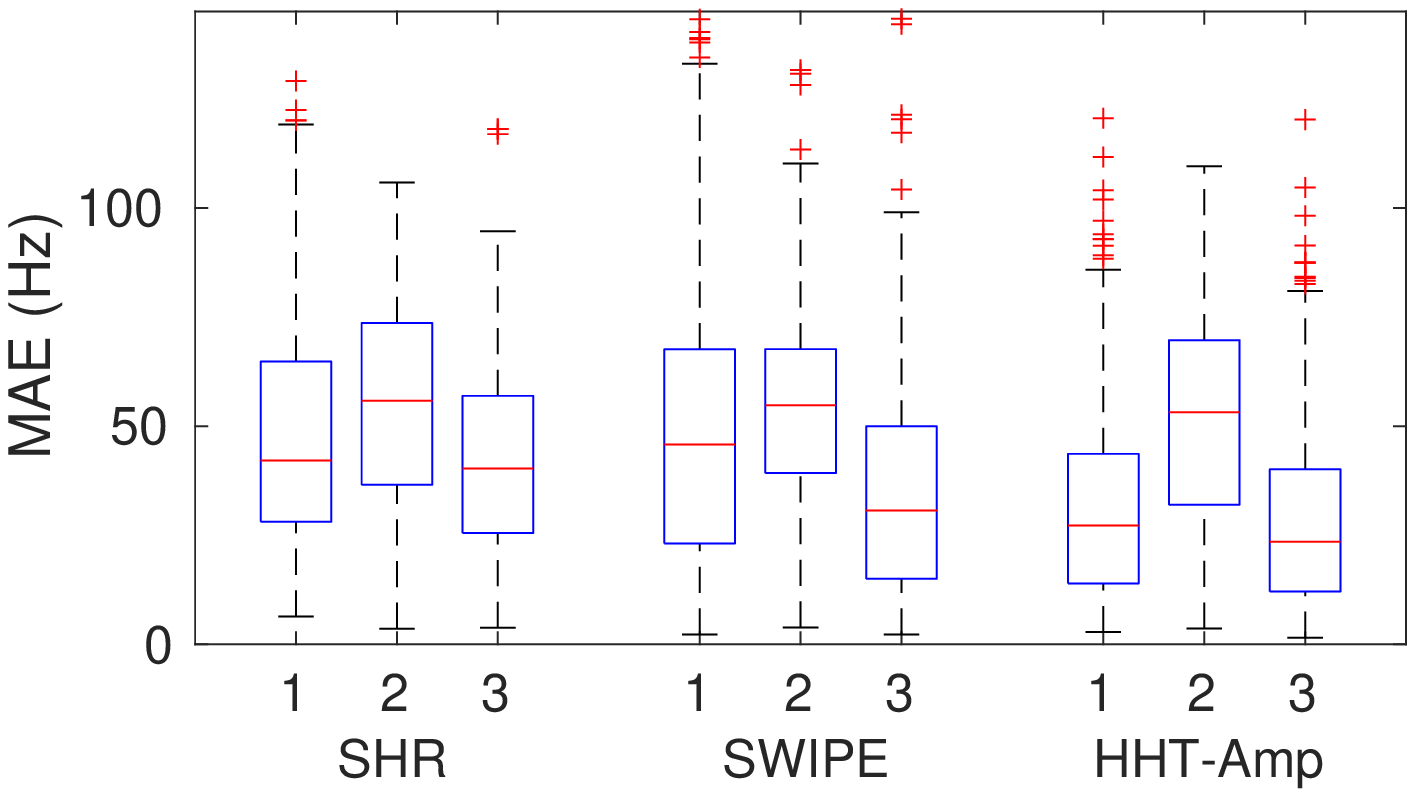}
}
\hspace{-0.3cm}
 \subfigure[]{
  \includegraphics[width=0.31\linewidth,keepaspectratio=true,clip=true,trim=0pt 0pt 0pt 15pt]{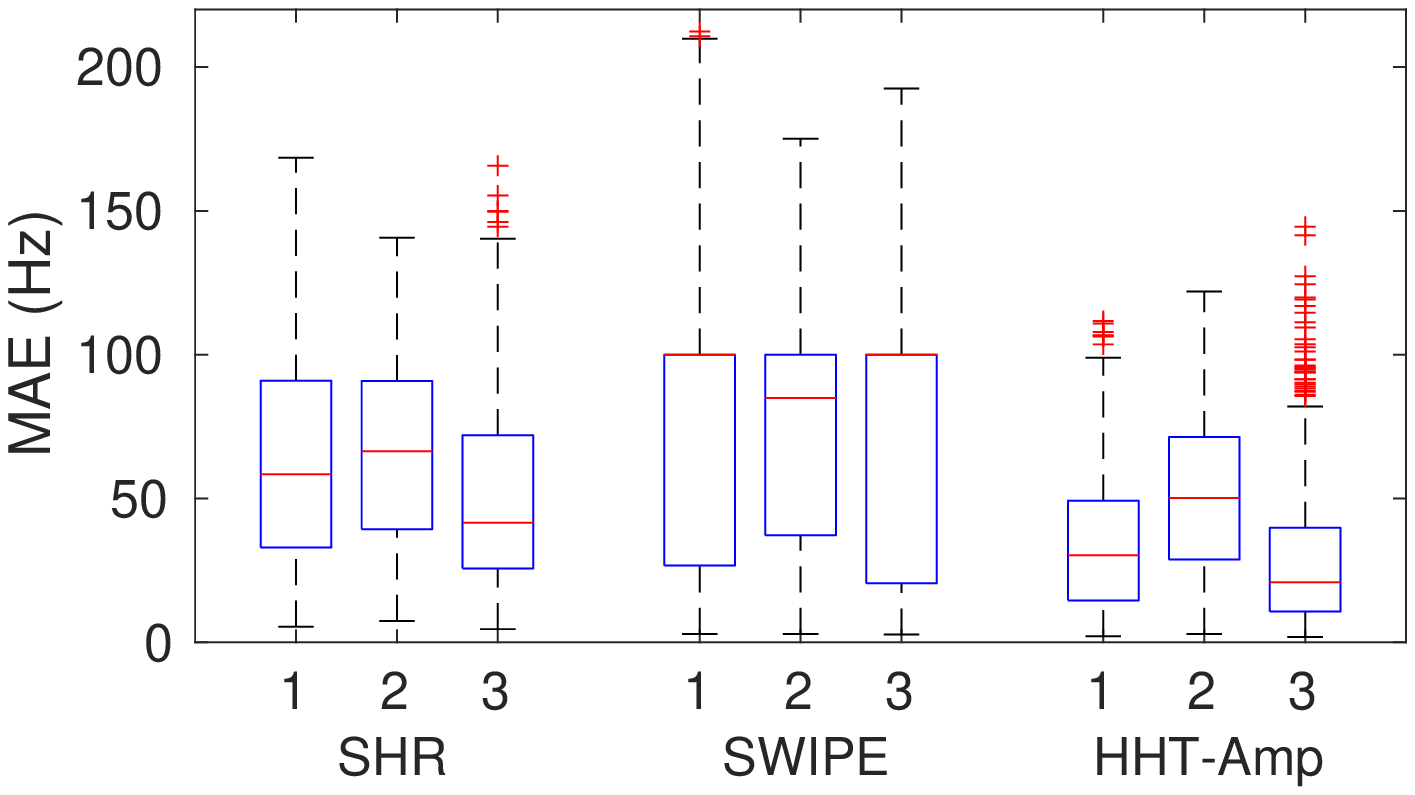}
}
\hspace{-0.3cm}
  \subfigure[]{
  \includegraphics[width=0.31\linewidth,keepaspectratio=true,clip=true,trim=0pt 0pt 0pt 15pt]{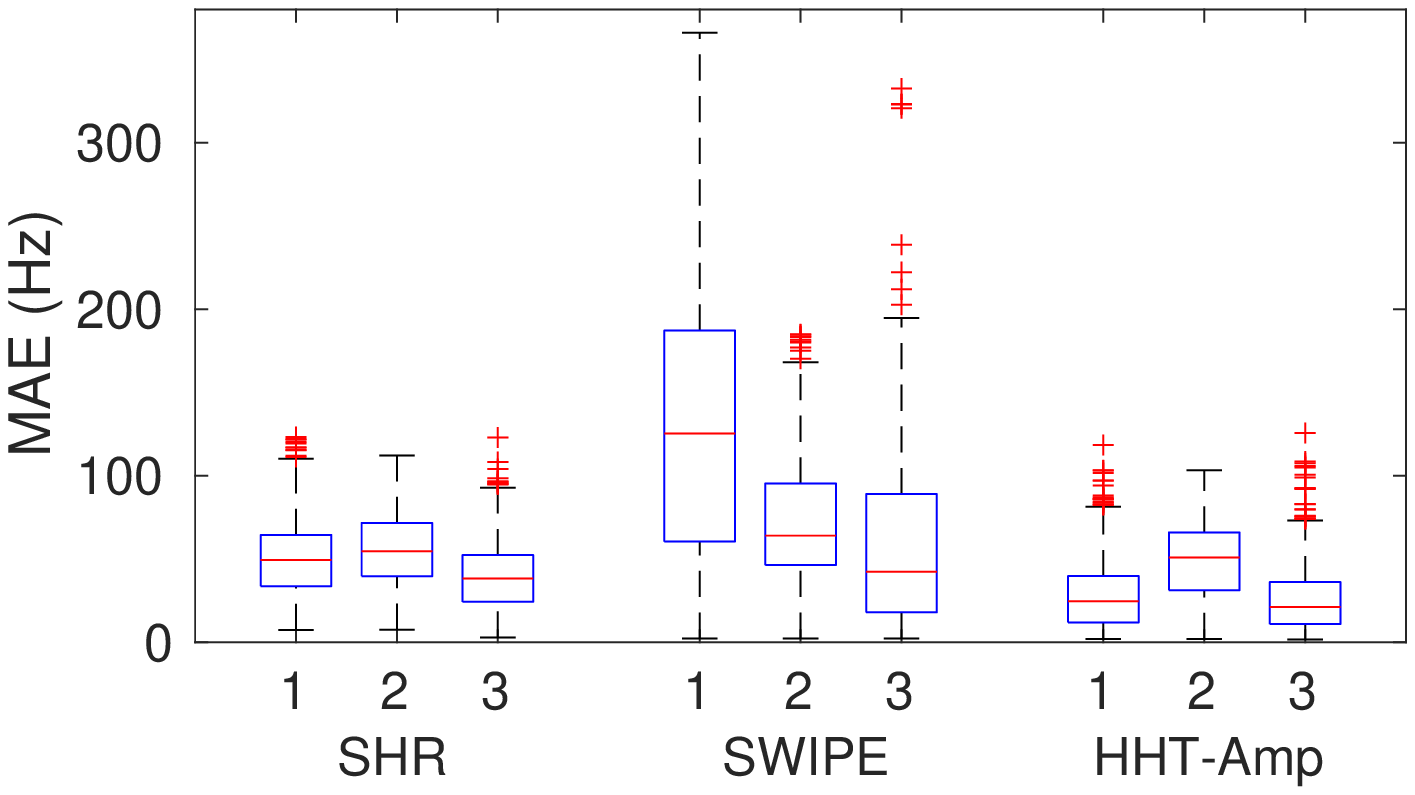}
}
\hspace{-0.01cm}
 \subfigure[]{
  \includegraphics[width=0.31\linewidth,keepaspectratio=true,clip=true,trim=0pt 0pt 0pt 15pt]{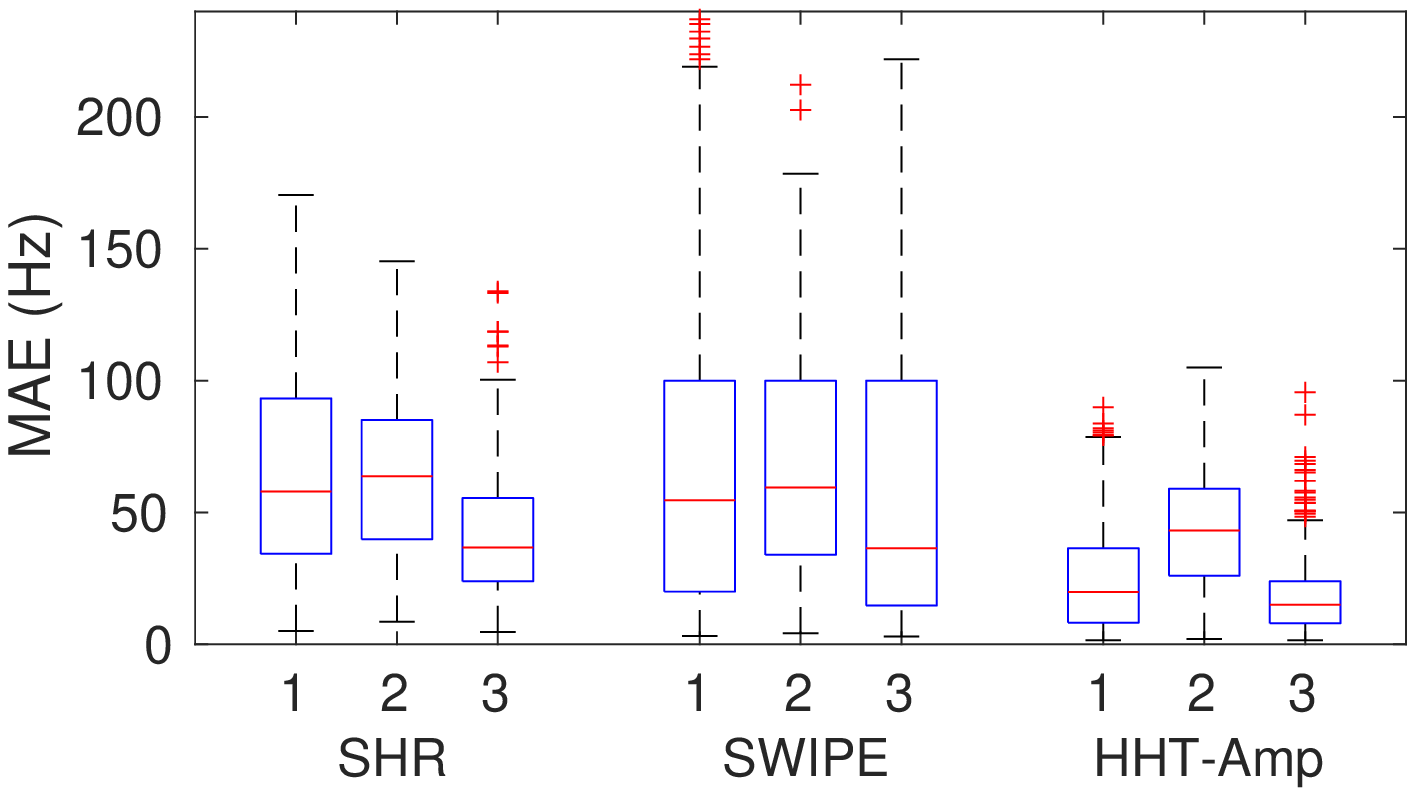}
}
\hspace{-0.3cm}
 \subfigure[]{
  \includegraphics[width=0.31\linewidth,keepaspectratio=true,clip=true,trim=0pt 0pt 0pt 15pt]{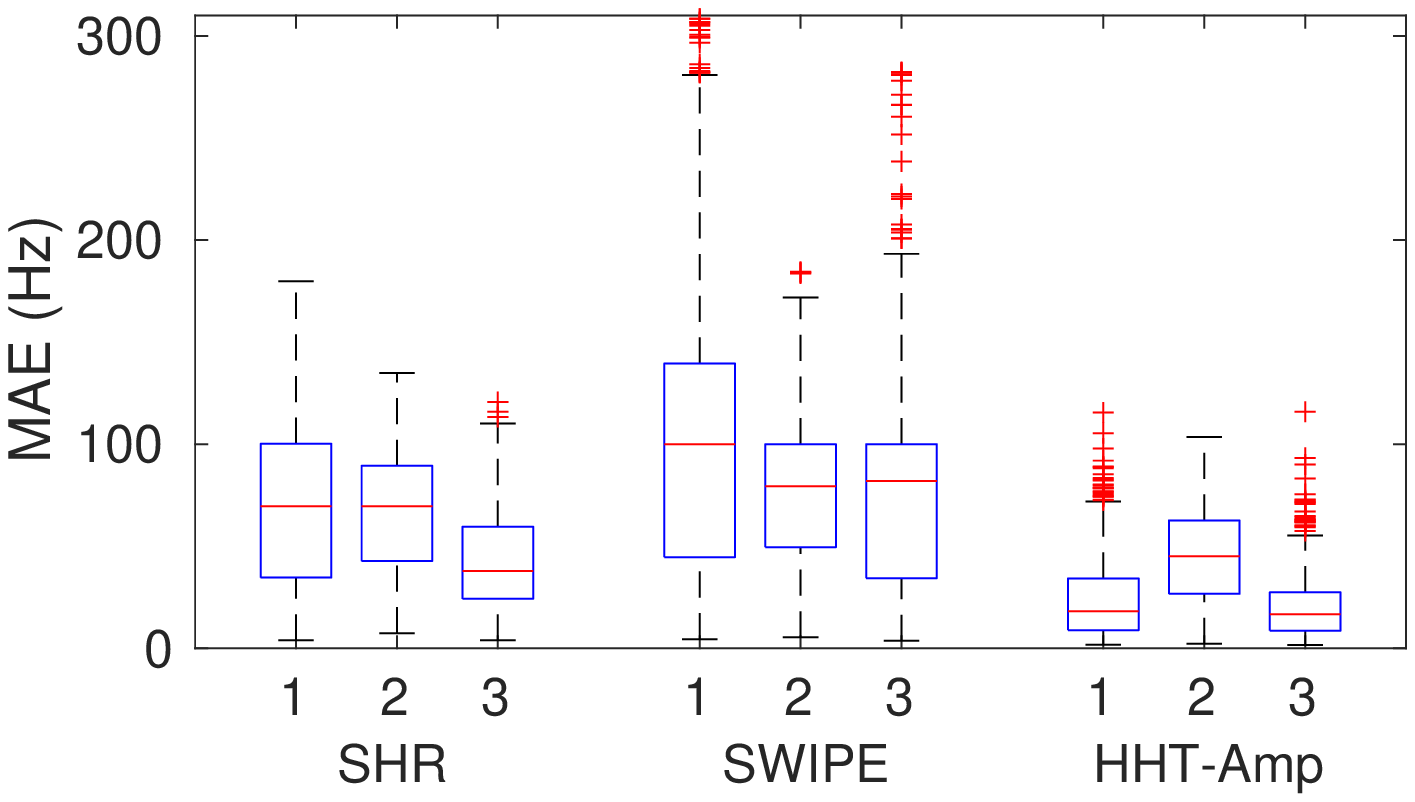}
}
\hspace{-0.3cm}
 \subfigure[]{
  \includegraphics[width=0.31\linewidth,keepaspectratio=true,clip=true,trim=0pt 0pt 0pt 15pt]{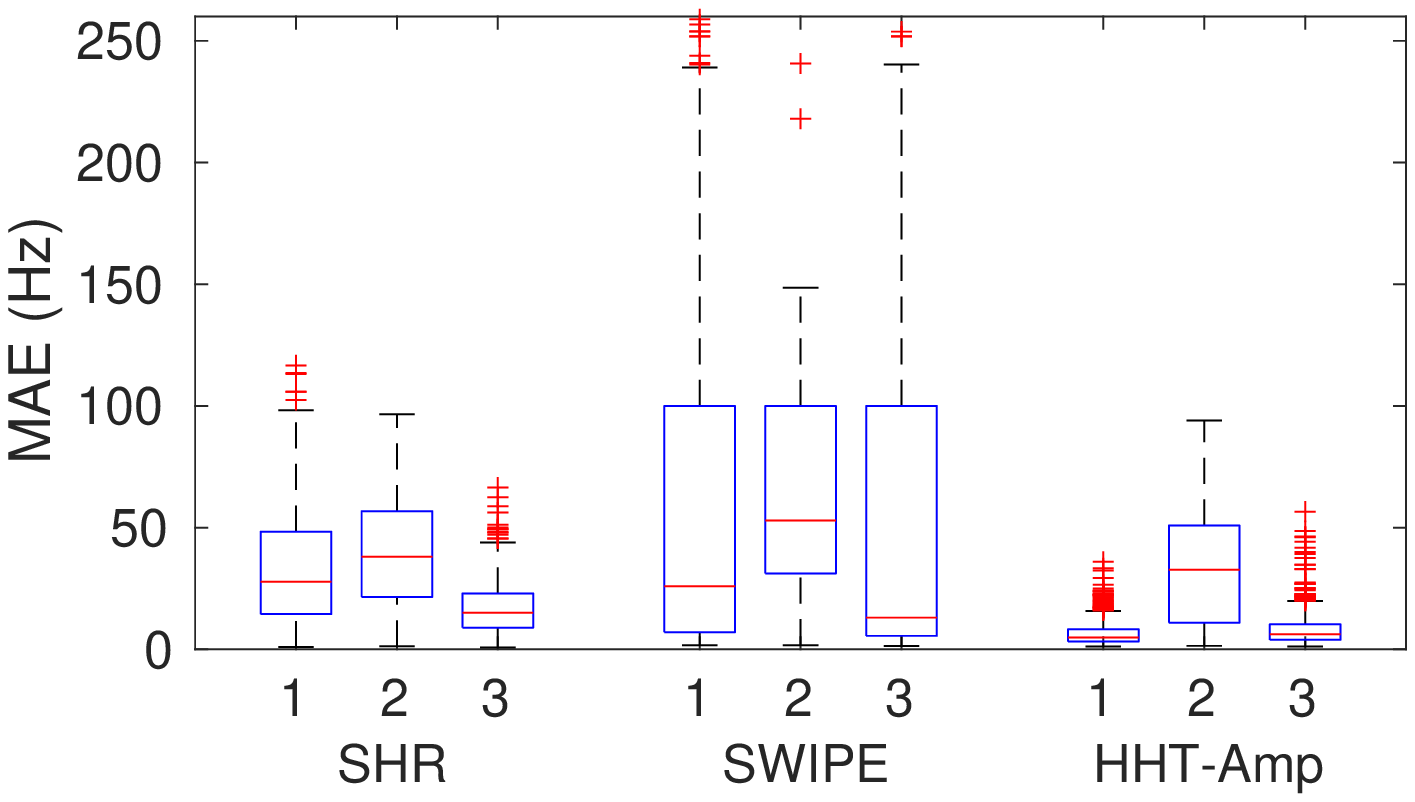}
}
 \caption{The MAE box-plots for TIMIT database speech signals with five SNR values, for six noisy conditions: (a) Babble, (b) SSN, (c) Cafeteria, (d) Train, (e) Helicopter, and (f) Volvo. Case 1 refers to original F0 estimation method, case 2 the improved F0 by DCNN and case 3 improved F0 by PRO. Results obtained considering the low/high frequency separation errors.}
\label{MAE_Ruidos_TIMIT}

\end{figure*}

Fig. \ref{MAE_Ruidos_CSTR} and Fig. \ref{MAE_Ruidos_TIMIT} illustrate the MAE results obtained  for six noisy conditions with the CSTR and TIMIT databases, respectively. Each box-plot denotes the MAE values achieved with five SNR values: -15 dB, -10 dB, -5 dB, 0 dB and 5 dB. Note that again the proposed method achieved the lowest MAE values. For CSTR database, the DCNN strategy attains improvement in original F0 estimation in some cases of SHR and SWIPE, particularly for SSN, Cafeteria, and Helicopter noises. The MAE values for the TIMIT database indicate that PRO improved the F0 estimation accuracy, overcoming the DCNN baseline solution. Once again, the PRO method leads the HHT-Amp estimator to achieve best accuracy results for both databases, with exception for Volvo noise with TIMIT (Fig. 6(f)).

In summary, the proposed method leads to the best results when compared to the solution DCNN, in terms of low/high frequency separation accuracy, and improvement of fundamental frequency estimation accuracy. In addition, the combination of PRO with the HHT-Amp algorithm (PRO + HHT-Amp) attained the lowest GE and MAE scores, for both the CSTR and TIMIT databases. Furthermore, SHR baseline approach are also outperformed by the HHT-Amp. The fact that SHR adopts two F0 candidates, against the three candidates considered in HHT-Amp, can favor this last one in the true F0 selection. In contrast, baseline techniques based on the SWIPE  estimation reach the highest error values. This occurs because SWIPE is very sensitive to the severe noise disturbances, which may cause expressive voiced/unvoiced detection errors. Finally, it is interesting to observe that even considering the classification errors in the estimation accuracy results, the proposed method improves F0 estimation of speech signals corrupted by different types of noise, and hence it can be used in real world applications.

\begin{table}[t!] \caption{\label{complexidade} Normalized Mean Processing Time.}

\renewcommand{\arraystretch}{1.3}
\setlength{\tabcolsep}{2.pt}
\begin{center}
{
\begin{tabular}{ccccccccccc}

\hline
\multicolumn{3}{c}{Original}&&\multicolumn{3}{c}{DCNN}&&\multicolumn{3}{c}{PRO}\\
\cline{1-3}\cline{5-7}\cline{9-11}
SHR&\scriptsize{SWIPE}&\scriptsize{HHT}\tiny{-Amp}&&SHR&\scriptsize{SWIPE}&\scriptsize{HHT}\tiny{-Amp}&&SHR&\scriptsize{SWIPE}&\scriptsize{HHT}\tiny{-Amp}\\\hline

0.03&0.03&0.93&&0.22&0.20&1.07&&0.94&0.95&1.00\\ \hline
\end{tabular}
}
\end{center}
\vspace{-0.3cm}
\end{table}

Table \ref{complexidade} indicates the computational complexity which refers to the normalized processing time required for each scheme evaluated for 512 samples per frame. These values are obtained with an Intel (R) Core (TM) i7-9700 CPU, 8 GB RAM, and are normalized by the execution time of the most accurate method PRO + HHT-Amp. The processing time required for the training process by DCNN is not considered here. Therefore, note that the HHT-Amp baseline scheme and the proposed method present a longer processing time, since they are based on the EEMD, and demands a relevant computational cost.

\section{Conclusion}

This paper introduced a method for low/high frequency separation of voiced frames, in order to improve the F0 estimation accuracy. The EEMD algorithm was applied to decompose the noisy speech signal. Then, F0 estimation and selection of analyzed decomposition modes was performed to detect the low or high frequency region of the voiced frames. According to this separation, the candidates of F0 estimation techniques were corrected, improving their accuracy. Several experiments were conducted to evaluate the improvement provided by the PRO method and the DCNN based approach, with three baseline F0 estimation techniques. Two speech databases and six acoustic noises of different sources were adopted for this purpose. Results were examined considering separation accuracy, and estimation error measures GE and MAE. The proposed method attained the smallest errors in low/high frequency separation in all cases, when compared to DCNN. Furthermore, the F0 estimation metrics demonstrated that PRO leads to superior improvement in F0 estimation accuracy. Particularly, the PRO + HHT-Amp method outperformed the baseline approaches in terms of GE and MAE, with interesting accuracy in F0 detection in various noisy environments. Future research includes the investigation of the low/high frequency separation in other tasks, such as for intelligibility improvement.

\ifCLASSOPTIONcaptionsoff
  \newpage
\fi



%




\bibliographystyle{ieeetr}
\bibliography{bare_jrnl}

%

%
%




\end{document}